\documentclass[preprint,12pt]{elsarticle}

\usepackage{subfig}
\usepackage{amssymb}
\usepackage{amsmath}
\usepackage{multirow}
\usepackage{graphicx}
\usepackage{pdfpages}
\usepackage{epstopdf}
\usepackage[makeroom]{cancel} 
\usepackage{xcolor}
\usepackage{float}
\usepackage[margin=1in]{geometry}
\usepackage{setspace}

\pdfoutput=1

\journal{Industrial \& Engineering Chemistry Research}

\begin{document}

\def\eq#1{Eq.~(\ref{#1})}
\newcommand{\p}{\partial}
\newcommand{\rd}{{\rm d}}
\newcommand{\ra}{{\rm a}}
\newcommand{\rb}{{\rm b}}
\newcommand{\rRe}{{\rm Re}}
\newcommand{\rD}{{\rm D}}
\newcommand{\dpm}{{\rm dpm}}
\newcommand{\rr}{{\rm r}}
\newcommand{\rL}{{\rm L}}
\newcommand{\rg}{{\rm g}}
\newcommand{\rk}{{\rm k}}
\newcommand{\lift}{{\rm lift}}
\newcommand{\wall}{{\rm wall}}
\newcommand{\rt}{{\rm t}}
\newcommand{\rx}{{\rm x}}
\newcommand{\rrho}{{\rm  rgh}}
\newcommand{\rmax}{{\rm  max}}
\newcommand{\rmin}{{\rm  min}}
\newcommand{\turb}{{\rm  turb}}
\newcommand{\rp}{{\rm p}}
\newcommand{\rAU}{{\rm rAU}}
\newcommand{\kd}{{\rm Kd}}
\newcommand{\vm}{{\rm Vm}}
\newcommand{\eff}{{\rm eff}}
\newcommand{\rc}{{\rm c}}
\newcommand{\nablaT}{{\nabla ^\mathrm{T}}}
\newcommand{\m}{{\rm m}}
\newcommand{\rT}{{\rm T}}
\newcommand{\bfq}{{\bf q}}
\newcommand{\bfn}{{\bf n}}
\newcommand{\rP}{{\rm P}}
\newcommand{\rgh}{{\rm rgh}}
\newcommand{\Sp}{{\rm Sp}}
\newcommand{\Su}{{\rm Su}}
\newcommand{\rN}{{\rm N}}
\newcommand{\bfU}{\mathbf{U}}
\newcommand{\bfS}{\mathbf{S}}
\newcommand{\bfh}{\mathbf{h}}
\newcommand{\HbyA}{\mathbf{HbyA}}
\newcommand{\bfI}{\mathbf{I}}
\newcommand{\bfP}{\mathbf{P}}
\newcommand{\bfR}{\mathbf{R}}
\newcommand{\bfx}{\mathbf{x}}
\newcommand{\bfA}{\mathbf{A}}
\newcommand{\bftau}{\mathbf{\tau}}
\newcommand{\bfM}{\mathbf{M}}
\newcommand{\bfK}{\mathbf{K}}
\newcommand{\bfH}{\mathbf{H}}
\newcommand{\bfg}{\mathbf{g}}
\newcommand{\bfb}{\mathbf{b}}
\newcommand{\bfy}{\mathbf{y}}
\newcommand{\bfD}{\mathbf{D}}
\newcommand{\bfd}{\mathbf{d}}
\newcommand{\sd}{\, {\rm d}}

\begin{frontmatter}
\title{Investigation of the force closure for Eulerian-Eulerian simulations: a validation study of nine gas-liquid flow cases}

\author[a,b]{Yulong Li}
\author[c,d]{Dongyue Li\corref{author2}}

\address[a]{School of Marine Engineering and Technology, Sun Yat-Sen University, Zhuhai, China}
\address[b]{Southern Marine Science and Engineering Guangdong Laboratory (Zhuhai), Zhuhai, China}
\address[c]{State Key Laboratory of Advanced Metallurgy, University of Science and Technology Beijing, Beijing, China}
\address[d]{DYFLUID Ltd, Beijing, China}
\cortext[author2] {Corresponding author. Email: li.dy@dyfluid.com, dongyueli@ustb.edu.cn (D. Li)}

\begin{abstract}
Gas-liquid flows can be simulated by the Eulerian-Eulerian (E-E) method. Whether to include a specific momentum interfacial exchange force model remains as a question with no answer. In this work, our aim is to seek a general numerical settings for the E-E method, which can provide competent results for industrial bubbly flows with different geometries under different operations. They were selected from different industries including chemical, nuclear, bio-processing and metallurgical engineering.  Simulations were launched by the OpenFOAM solver \verb+reactingTwoPhaseEulerFoam+, in which the E-E method was implemented with sophisticated numerical techniques to ensure stabilities. Predictions were compared against experimental data. It was found that the drag force and turbulent dispersion force play the most important role on the predictions and should be included for all simulations. The first one accounts for the two-way coupling while the second one accounts for the turbulence effect and ensures the E-E equations to be well-posed. The lift force and wall lubrication force should be included to address the phase fraction accumulation in the vicinity of the wall, especially for pipe flows with large aspect ratio. In other cases the lateral forces can be safely neglected. All the test case are open-sourced and are available as supplementary data for anyone to download as baseline test cases. 
\end{abstract}

\begin{keyword}
Gas-liquid flows \sep Eulerian-Eulerian method \sep  Momentum interfacial exchange force  \sep OpenFOAM  
\end{keyword}

\end{frontmatter}

\section{Introduction}

Gas-liquid flows are encountered in a variety of applications and can be simulated by different models, such as the Eulerian-Eulerian (E-E) method, the Eulerian-Lagrangian (E-L) method and the direct method which is also called multi-phase direct numerical simulation (DNS). In the E-E method \cite{jakobsen2005modeling,
buwa2006eulerian,
li2009cfd,
li2009numerical,
besagni2017computational,
banaei2018tracking,
banaei2018particle,besagni2019prediction,
shi2019modelling}, each fluid phase is considered as a continuum in the computational domain under consideration which can interpenetrate with the other fluid phase. Several averaging methods (such as volume or ensemble averaging) are used to formulate basic governing equations. In the E-L approach \cite{buwa2006eulerian,
padding2015euler,
quiyoom2017euler,
battistella2018euler}, the continuous phase is treated in an Eulerian framework whereas the motion of individual bubbles is simulated by solving the force balance equation. The trajectories of these bubbles are computed in the control volume and averaged at the computational level. In the multi-phase DNS \cite{deen2014direct,
das2017dns,rabha2010volume,
goel2008numerical}, one employs the Navier-Stokes equations directly, without further manipulation and the topology of the interface between the two-phases is determined as part of the solution. No additional modelling assumptions are introduced.  DNS requires very high resolution in order to resolve a broad range of temporal and spatial scales.  These scales are associated with the topology of the interface, e.g. the size of the dispersed particles, or with the fluid motion, e.g. the eddies encountered in the turbulent motion. Resolving these scales is computationally expensive both in terms of computer memory size execution time. Therefore, DNS is
restricted to only low Reynolds numbers and a few bubbles/particles due to its high computational cost.

Thanks to the computational economics, many works used the E-E method to simulate gas-liquid flows. However, as a macro-scopic method, the E-E method 1) requires constitutive models (e.g., the solid-phase stress model); 2) requires models to address the momentum and energy exchange terms (e.g., the drag model); 3) loses the characteristic feature for those scales which is much smaller than the mesh resolution.  To obtain reasonable results, suitable sub-models and parameters should be adjusted. Moreover, the observed flow field is also different for the gas-liquid flows in different industries. For example, bubble columns are quite common in chemical engineering industry. The liquid flow field in the bubble column tends to form a circulation model due to the movement of the bubbles, which is commonly referred as ``Gulf-stream'' or ``cooling tower'' flow regime.  In the nuclear engineering, the phase fraction of the bubbles in vertical upward pipe shows a wall peak  due to the lateral movement of the small bubbles \cite{rzehak2017unified,
ziegenhein2019critical,
hessenkemper2019contamination}. Without loss of the generality, it was proven that the E-E method is able to capture the typical flow field information. However, amount of work needs to be studied for the sub-models, especially for the momentum interfacial exchange models.

The drag force and the buoyancy force, as the most important momentum interfacial exchange forces, determine the bubble terminal velocity and address the coupling between the disperse phase with the continuous phase. The lift force and wall lubrication force address the lateral movements of the bubbles. A correct description of the lift coefficient and wall force coefficient is crucial in order to model this transversal force correctly. The turbulence dispersion force acts as a driving force for bubbles to move from areas with higher phase fractions to areas of lower phase fraction. It arises due to the pressure variations in the continuous phase that are not resolved at meso-scale. It is also shown by Panicker et al. \cite{panicker2018hyperbolicity} and Vaidheeswaran and Lopez de Bertodano \cite{vaidheeswaran2017stability} that the addition of a dispersion term ensures the hyperbolicity of the PDE, and prevents the non-physical instabilities in the predicted multiphase flows upon grid refinement.

Although there is universal consensus in the literature on the need to incorporate drag into any model of a bubble column, many works admit that there is still no agreement in the community on the other forces to be used at best \cite{ma2016large}. The purpose of the present contribution is to simulate large amount of gas-liquid flows with different force closure combinations to investigate their importance. Our aim is to seek a general numerical settings for the E-E method, which can provide competent results for industrial bubbly flows. 

\section{Description of models}
Within the Eulerian framework, two sets of Navier-Stokes equations are ensemble-averaged, and the effects of turbulence and inter-phase phenomena are taken into account using closure models. The conservation of mass for phase $a$ and phase $b$ can be expressed by \cite{drew1982mathematical}
:\begin{equation}\label{contAlpha}
\frac{{\p \left( {{\alpha _\ra }{\rho_\ra }} \right)}}{{\p t}} + \nabla  \cdot \left( {{\alpha_\ra }{\rho_\ra}{\bfU_\ra}} \right) = 0,
\end{equation}
\begin{equation}\label{contAlpha2}
\frac{{\p \left( {{\alpha _\rb}{\rho_\rb}} \right)}}{{\p t}} + \nabla  \cdot \left( {{\alpha_\rb}{\rho_\rb}{\bfU_\rb}} \right) = 0,
\end{equation}
where $\alpha_\ra$ and $\alpha_\rb$ are the phase fraction of phase $a$ and phase $b$, $\rho_\ra$ and $\rho_\rb$ are the density for phase $a$ and phase $b$, and $\bfU_\ra$ and $\bfU_\rb$ are the average velocity for phase $a$ and phase $b$, respectively.  The average velocities $\bfU_\ra$ and $\bfU_\rb$ can be calculated by solving the corresponding phase momentum equations \citep{drew1982mathematical}:
\begin{equation}\label{momentum}
\frac{{\p \left( {{\alpha_\ra }{\rho_\ra }{\bfU_\ra }} \right)}}{{\p t}} + \nabla \cdot \left( {{\alpha_\ra}{\rho_\ra } {{\bfU_\ra} {\bfU_\ra}} } \right) - \\ \nabla  \cdot \left( {{\alpha_\ra}\rho_\ra{\bftau_\ra}} \right) 
= 
- {\alpha_\ra} \nabla p_\ra + {\alpha_\ra}{\rho_\ra} \bfg - {\bfM},
\end{equation}
\begin{equation}\label{momentum2}
\frac{{\p \left( {{\alpha_\rb}{\rho_\rb}{\bfU_\rb}} \right)}}{{\p t}} + \nabla \cdot \left( {{\alpha_\rb}{\rho_\rb} {{\bfU_\rb}  {\bfU_\rb}} } \right) - \\ \nabla  \cdot \left( {{\alpha_\rb}\rho_\rb{\bftau_\rb}} \right)  
= 
- {\alpha_\rb} \nabla p_\rb + {\alpha_\rb}{\rho_\rb} \bfg + {\bfM},
\end{equation}
where $p_\ra$ and $p_\rb$ are the pressure for each phase, $\bftau_\ra$ and $\bftau_\rb$ are the effective Reynolds stress tensors, $\bfg$ is the gravitational acceleration vector, and $\bfM$ is the interfacial force exchange term. It is common to break it down into the axial drag force $\bfM_{\mathrm{drag}}$; the lateral lift force $\bfM_{\mathrm{lift}}$, which acts perpendicular to the direction of the relative motion of the two phases \cite{tomiyama2002transverse}; the wall lubrication force $\bfM_{\mathrm{wall}}$ which acts to drive the bubble away from the wall \cite{antal1991analysis}; and the turbulent dispersion force $\bfM_{\mathrm{turb}}$, which is the result of the turbulent fluctuations of the liquid velocity \cite{lopez1993turbulent}. Specifically, the drag force can be calculated as follows:
\begin{equation}\label{drag}
\bfM_{\mathrm{drag}}=\frac{3}{4}\alpha_\ra\rho_\rb C_\rD\frac{1}{d_\ra} \left|\bfU_\rb-\bfU_\ra\right| \left(\bfU_\rb-\bfU_\ra\right),
\end{equation}
where $C_\mathrm{D}$ is the drag force coefficient and $d_\ra$ is the diameter of phase $a$. The lift force can be calculated as follows \cite{tomiyama1995effects}:
\begin{equation}\label{lift}
\bfM_\lift=\alpha_\rb C_\rL\rho_\ra\bfU_\rr\times\left(\nabla\times\bfU_\rb\right),
\end{equation}
where $C_\mathrm{L}$ is the lift force coefficient, $\bfU_\rr$ is the relative velocity which equals $\bfU_\rb-\bfU_\ra$. The wall lubrication force can be calculated as follows \cite{antal1991analysis}:
\begin{equation}\label{wall}
\bfM_\wall=C_\wall\rho_\rb\alpha_\ra|\bfU_\rr|^2\cdot\bfn,
\end{equation}
where $C_\wall$ is the wall lubrication force coefficient. The turbulence dispersion force can be calculated as follows \cite{lopez1993turbulent}:
\begin{equation}\label{turb}
\bfM_\turb=C_\rT\rho_\rb k_\rb\nabla\alpha_\ra,
\end{equation}
where $C_\rT$ is the turbulence dispersion force coefficient.  A comprehensive discussion of the E-E method can be found in other latest work \cite{ishii2010thermo,de2016two}. Readers are referred to the our previous work \cite{li2017simulation} and the Appendix for the finite volume discretization of the E-E method. 

\begin{figure}[H]
\begin{center}
\includegraphics
[width=0.4\textwidth]{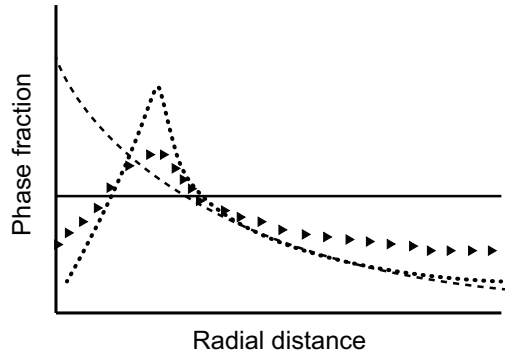}
\end{center}
\caption{Typical phase fraction profile predicted by CFD for
small bubbles (e.g., $d < 5$ mm) in vertical upward pipe. Solid line: Only drag force. Dashed line: Drag + lift force. Dots: Drag + lift + wall
lubrication force. Triangles: Drag + lift + wall lubrication + turbulent
dispersion force. } 
\label{fig1}
\end{figure} 

In the E-E method the interaction between the bubbles and the liquid phase is modelled through the momentum interfacial exchange forces in the momentum equation of the liquid and the gas phase. There is still no agreement in the community on the closures to be used at best. Physically, the effect of these forces are reported in Fig. \ref{fig1}. In a quiescent environmental, the buoyancy force pushes the bubbles upwardly. On the other hand, the drag force tends to exert on the opposite direction of the bubble movement. The direction of the drag force is equivalent to the direction of relative velocity. In the simplest cases, these two forces can reach to a balance and the terminal bubble velocity can be determined. 

The lift force arises due to the presence of shear in the liquid phase and acts perpendicular to the bubble rise direction. Its direction depends on the value of sign of $C_\rL$. Different models can be used, such as the constant lift coefficient model and the model developed by Tomiyama \cite{tomiyama2002transverse}. Experiments show that the movement of large bubbles and small bubbles is opposite, especially in pesudo-steady-state pipe flows. It can only be predicted by employing a non-constant lift coefficient. Thus, the lift force coefficient developed by Tomiyama is more reasonable. Moreover, the largest shear exists not only in the vicinity of no-slip walls, it may be also large when one phase was injected into another phase with high velocity. The wall lubrication force exists only in the vicinity of solid objects. It pushes the bubbles away from the wall avoiding bubble accumulation which is not observed from the experiments. It should be noted that this force exists only near the walls. Thus, the distance between the bubble and the nearest wall should be calculated in the wall force model. The turbulent dispersion force acts on the gradient of the phase fraction. It can be seen as a diffusion and it accounts for the random bubble movement in turbulent flows. It was found in many works that the results predicted by the E-E method is highly mesh depended and it was due to the mathmatical characteristic of the E-E equations \cite{vaidheeswaran2017stability,panicker2018hyperbolicity}. Including the turbulent dispersion force avoids the ill-posed problems and improves the stability of the E-E model. 

A large amount of force models were developed in the literature and it is not possible to compare all of them for all the test cases investigated in this work. Therefore, unless stated on intention, we employ the classic drag model by Ishii and Zuber \cite{ishii1979drag}, the lift force model by Tomiyama et al. \cite{tomiyama2002transverse}, the wall force by Antal et al. \cite{antal1991analysis} and the turbulent dispersion force by Loptez De Bertodano \cite{lopez1993turbulent} in our simulations. Since we focus on the validation between numerical results and experiments, readers are suggested to Appendix for the discretization of the governing equations.

\section{Test cases}

\begin{figure}[H]
\begin{center}
\subfloat[Test case A.1]
{\includegraphics
[width=0.33\textwidth]{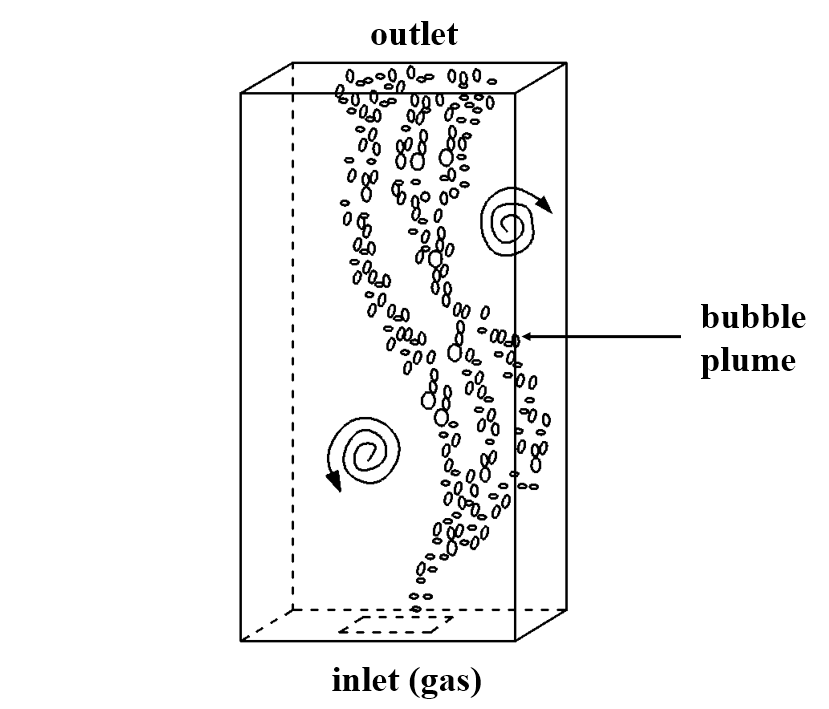}}
\subfloat[Test case A.2]
{\includegraphics
[width=0.33\textwidth]{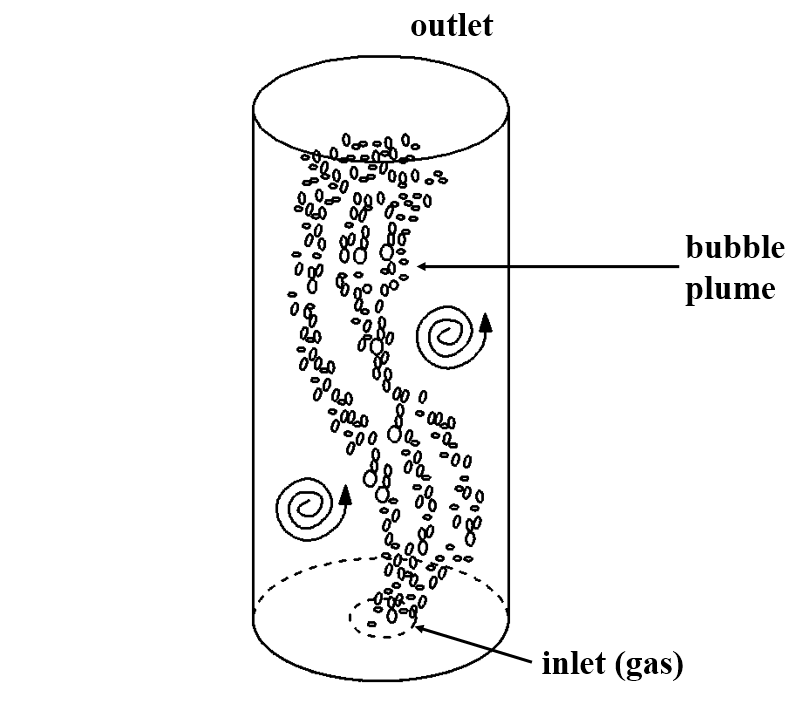}}
\subfloat[Test case A.3]
{\includegraphics
[width=0.33\textwidth]{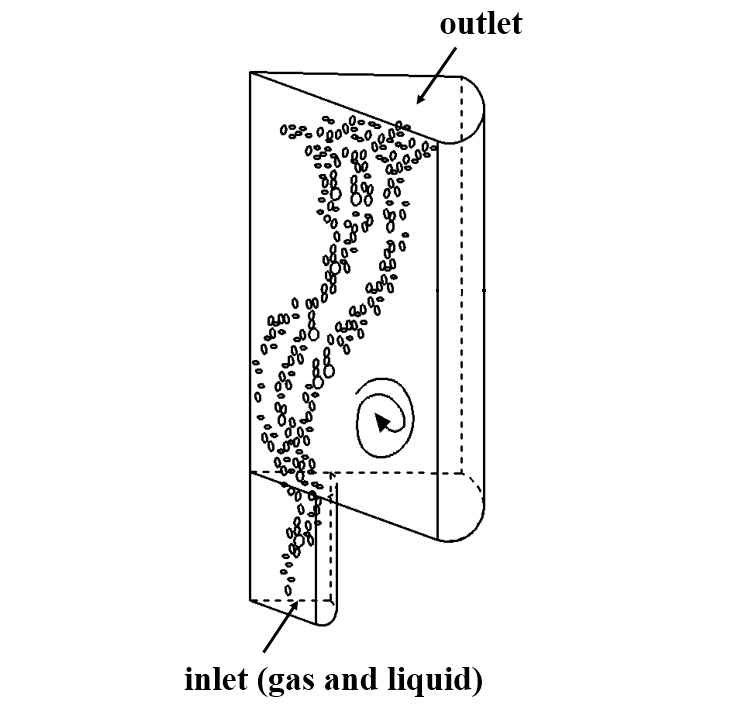}}

\subfloat[Test case B.1]
{\includegraphics
[width=0.33\textwidth]{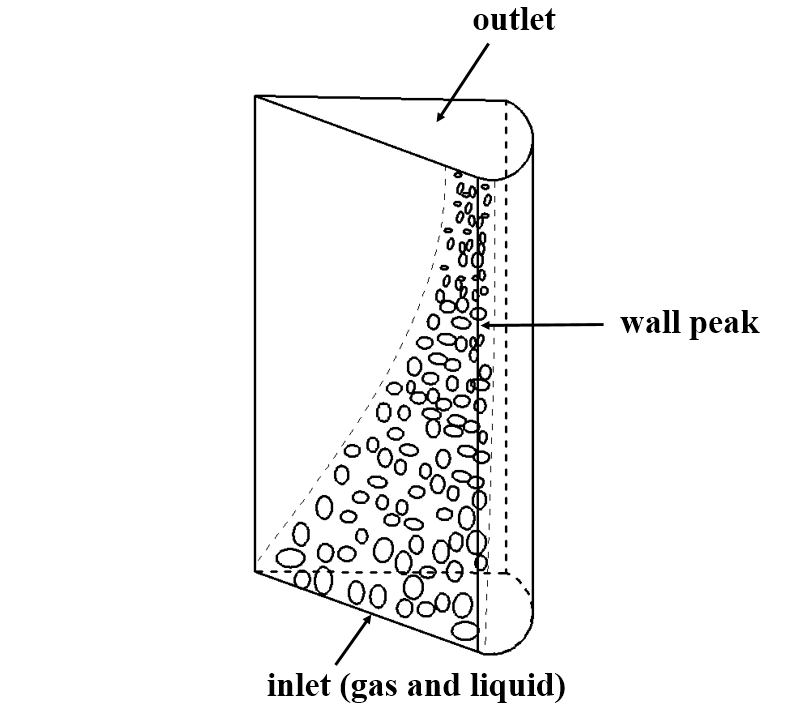}}
\subfloat[Test case B.2]
{\includegraphics
[width=0.35\textwidth]{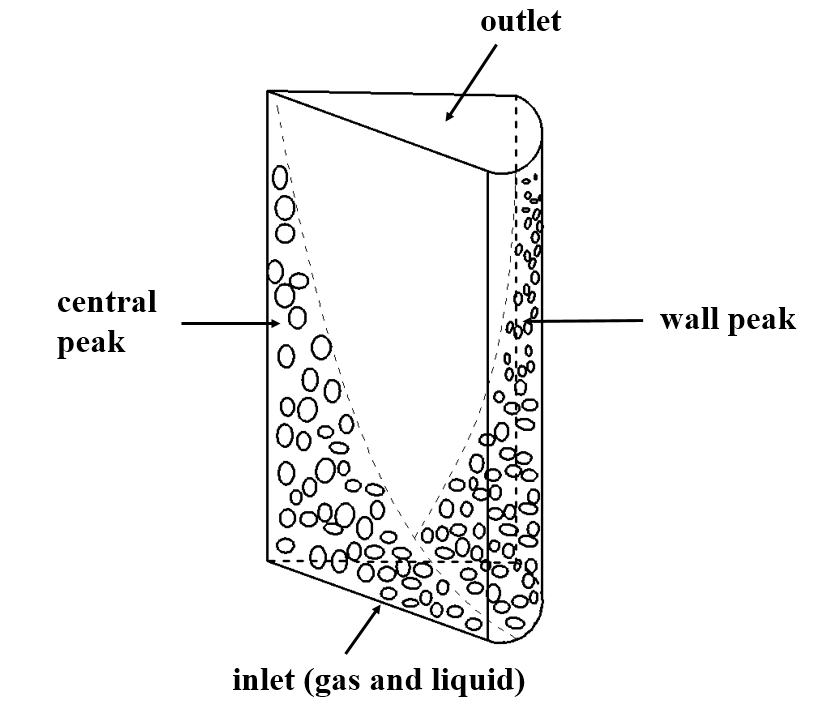}}
\subfloat[Test case B.3]
{\includegraphics
[width=0.28\textwidth]{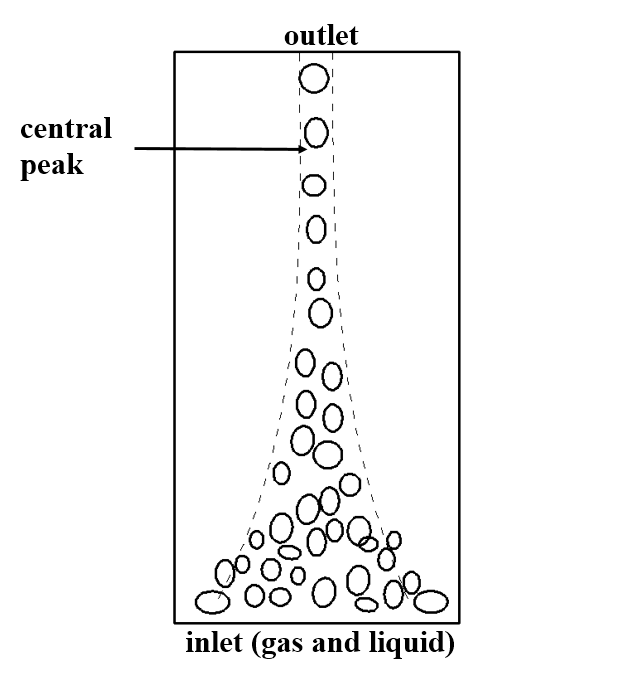}}

\subfloat[Test case B.4]
{\includegraphics
[width=0.33\textwidth]{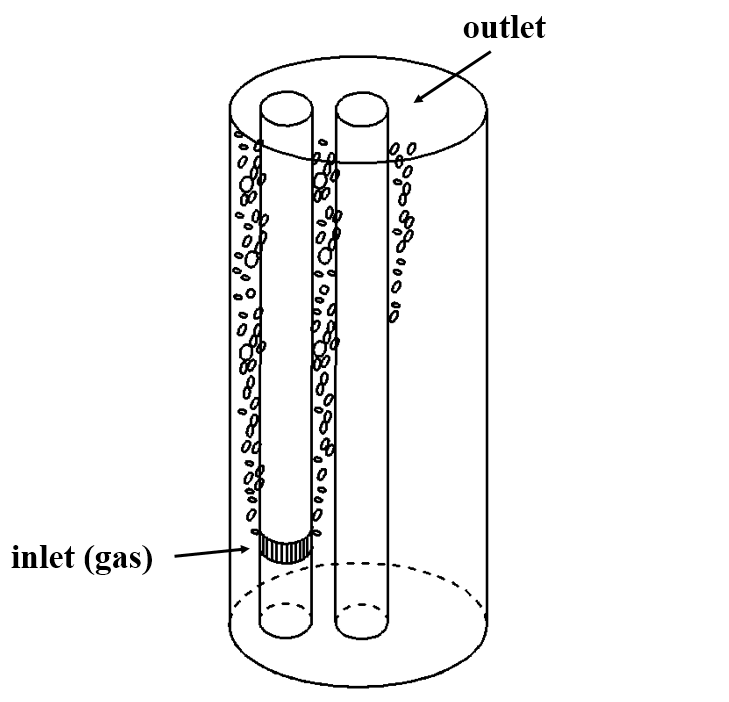}}
\subfloat[Test case C.1]
{\includegraphics
[width=0.33\textwidth]{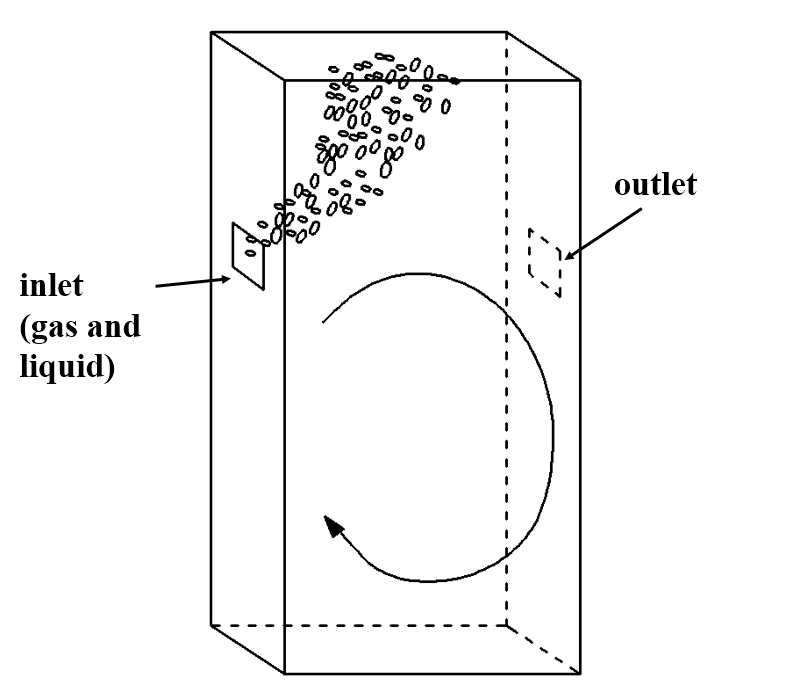}}
\subfloat[Test case C.2]
{\includegraphics
[width=0.33\textwidth]{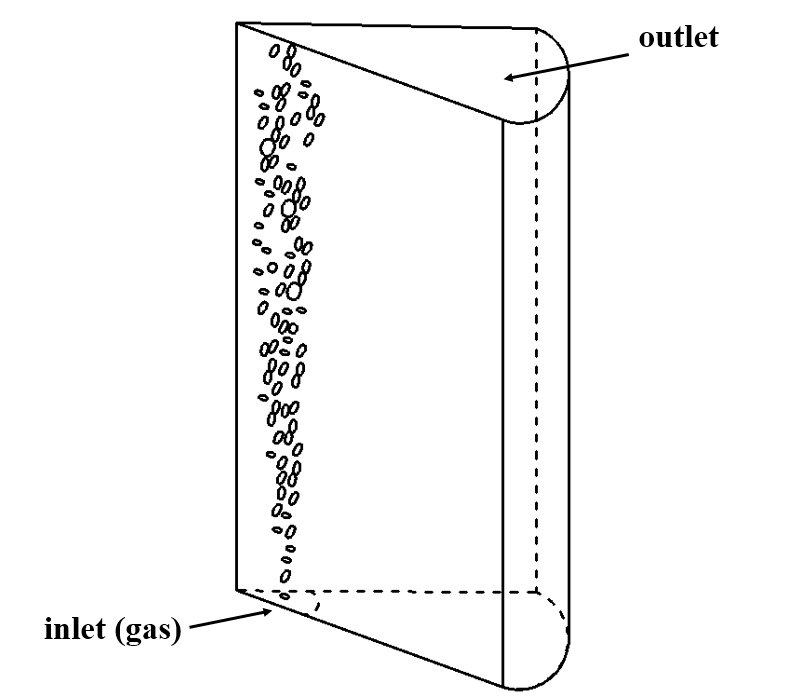}}

\end{center}
\caption{Schematic of the gas-liquid flows  investigated in this work. } 
\label{Geo}
\end{figure}

As previously mentioned, the lack of an ideal experiment, with a good mix of local and global measurements performed under a wide spectrum of operating conditions, necessary for a complete model validation, brought us to consider many different test cases with different geometries. Only by using a large amount of test cases can we find a general numerical settings. These test cases were selected from different works. The schematic representation of these test cases is reported in Fig. \ref{Geo}. All of them were investigated by experiments which implies they are suitable benchmark test cases for numerical investigations. Moreover, they were observed with different features as listed as follows:

\begin{itemize}

\item Test case A.1: investigation of bubble plumes in a rectangular bubble column of  \citet{diaz2008numerical}. The experimental equipment consists of a 0.2 m wide, 1.8 m high and 0.04 m deep partially aerated bubble column, filled with tap water up to 0.45 m from the bottom at room temperature and atmospheric pressure, while the air is fed through a sparger composed of eight centered holes of 1 mm of diameter and 6 mm pitch. This test case was proven to be a very interesting test case because the liquid vortices generated by the bubble plumes are a favorable factor for mixing and, therefore, for speeding up all transport processes \cite{sokolichin1997dynamic}. Additionally, the existence of flow structures showing unsteady liquid recirculation is a typical phenomenon in industrial-scale bubble columns. 
	
\item Test case A.2: investigation of gas-liquid flows in an industrial bubble column of  \citet{mcclure2014validation}. It is a partially aerated cylindrical bubble column of 0.19 m diameter with a multi-point sparger filled with water up to 1 m from the bottom. The aspect ratio ($L/D$) is about 5.  In the bio-processing industry, the bubble column height to diameter typically ranges from 2 to 5 and is usually operated in the heterogeneous flow regime to speed up mass and heat transfer. 
		
\item Test case A.3: investigation of sudden enlargement pipe flow of \citet{fdhila1991analyse}. The test case studied here is that of a bubbly air/water upward flow through a pipe with a sudden enlargement. The diameters of the two pipe sections are
50 mm and 100 mm, respectively.  The inlet phase fraction is characterised by a wall peak in experiments. 
The feature of this test case is that there is a large separation zone at the bottom of the pipe. This case has been  employed extensively to verify the E-E method and implementation  \cite{behzadi2004modelling,
oliveira2003numerical,
cokljat2006reynolds,
ullrich2017second}.

\item Test case B.1: investigation of bubbly flow in a cyliderial pipe of  \citet{lucas2005development}. Mixture of the gas and liquid is injected from the bottom pipe of 0.0256 m diameter and 3.53 m height. For the  vertical upward flow smaller bubbles tend to move towards the wall. A wall peak of the gas phase fraction occurs at the position of high $L/D$. This was also observed for single bubbles by  \citet{tomiyama1998drag}. In the case of vertical co-current pipe flow the radial flow field is symmetrically stable over a long distance. Therefore, this type of flow is well suited for the investigation of the non-drag forces. 
	
\item Test case B.2: investigation of bubbly flow in a circular pipe of  \citet{banowski2018experimental}. The experiment comprises a vertical pipe with 54.8 mm inner diameter and 6 m length. Gas and liquid mixture is injected from the bottom. It is similar with the test case B.2. However, besides the typical wall peak formed due to the smaller bubbles movement, a double peak of the phase fraction can also be observed due to the existence of large bubbles because the large bubbles tend to move to the center. 
	
\item Test case B.3: investigation of bubbly flow in a rectangle pipe of  \citet{vzun1990mechanism}. This test case is similar with test case B.3 with slightly difference of the geometry. Mixture of the gas and liquid is injected from the bottom of a rectangle channel of 0.0254 m length and 2 m height. Only the central peak of the gas phase fraction was observed.
	
\item Test case B.4: investigation of bubbly flow of \citet{besagni2016annular}. The gas-liquid flow  in an annular gap bubble column with two non-regular internal pipes was investigated. It consists of a non-pressized vertical column with an inner diameter 0.24 m and a height 5.3 m. In simulations the height of the domain is limited to 5 m. Two internal pipes are placed inside the column: one centrally positioned (with an external diameter of 0.06 m) and one asymmetrically positioned (with an external diameter of 0.075 m). The sparger is modeled as a uniform cylindrical surface with a height of 0.01 m placed on the lateral inner pipe at the vertical position of 0.3 m from the bottom of the domain. The aspect ratio of the geometry is small. Due to the existence of the non-irregular components, the gas phase fraction distribution developed to be quite flatten and no wall peak was observed in the experiments, even it can be seen as a pipe flow. 
	
\item Test case C.1: investigation of gas-liquid flows in a continuous casting molds of \citet{iguchi2000water}. The geometry employed in this test case is quite different with previous ones. The gas and liquid is injected into a rectangular vessel of 0.3 m length and 0.15 m width. In the experiments, it was observed that larger bubbles are lifted towards the liquid surface due to the buoyancy force acting on them, while smaller bubbles are carried deep. Such phenomenon is also known as phase segregation or poly-dispercity in other works, which was proven as a tough work for the E-E method. 
		
\item Test case C.2: investigation of gas-liquid flows in a continuous casting molds of \citet{sheng1993measurement}. The gas phase is injected from the bottom of the vessel of 0.76 m height and 0.5 m diameter. In this test case, the measured turbulence fields, gas phase fraction distribution, gas/liquid velocities in the plume zone were used for validation of various turbulence models. It can be seen as a suitable test case to validate the multi-phase turbulence model against experimental data. 
	
	
\end{itemize}

The reader may find these test cases are sorted by different groups. The bubble columns investigated in test case A.1 - A.3 are mainly encountered in chemical and bio-processing engineering. The aspect ratio of the geometry is small. They are usually used to speed up mixing and heat/mass transfer and are typical operated in medium or high superficial velocity. A strong coupling between the disperse phase and continuous phase is observed. The local phase fraction may be  high. Meanwhile, the liquid flow in these bubble columns may be highly transient and full of chaos with large-scale vortices. The pipe flow investigated in test case B.1 - B.4 are mainly encountered in nuclear engineering process. The aspect ratio of these pipes are relatively large. Different flow types exist depending on the gas flow rate. Dilute bubbly flows are found when operated at low gas superficial velocity. In bubbly flows, the liquid flow is rather stable and a steady-state can be achieved. The shape of the phase fraction distribution develops gradually to a stable distribution along the pipe axial direction, since the aspect ratio is quite large which implies the gas phase has enough time to develop. The bubbly flows investigated in test case C.1 - C.2 are mainly encountered in metallurgical engineering. The scale of the gas-liquid reactor is the same with that employed in the chemical engineering. The aspect ratio is also small. However, non-irregular design is quite common and the flow field is complex due to the existence of the complex geometry. Meanwhile, the small bubbles in the liquid phase, due to the phase segregation movement, are usually seen as impurity which need to be removed. 
At last, it should be stressed that although these test cases coming from different industries are full of different features, the core problem lies in the investigation of the gas-liquid flows by the numerical method.

\begin{figure}[H]
\begin{center}
\subfloat[Test case A.1]
{\includegraphics
[width=0.3\textwidth]{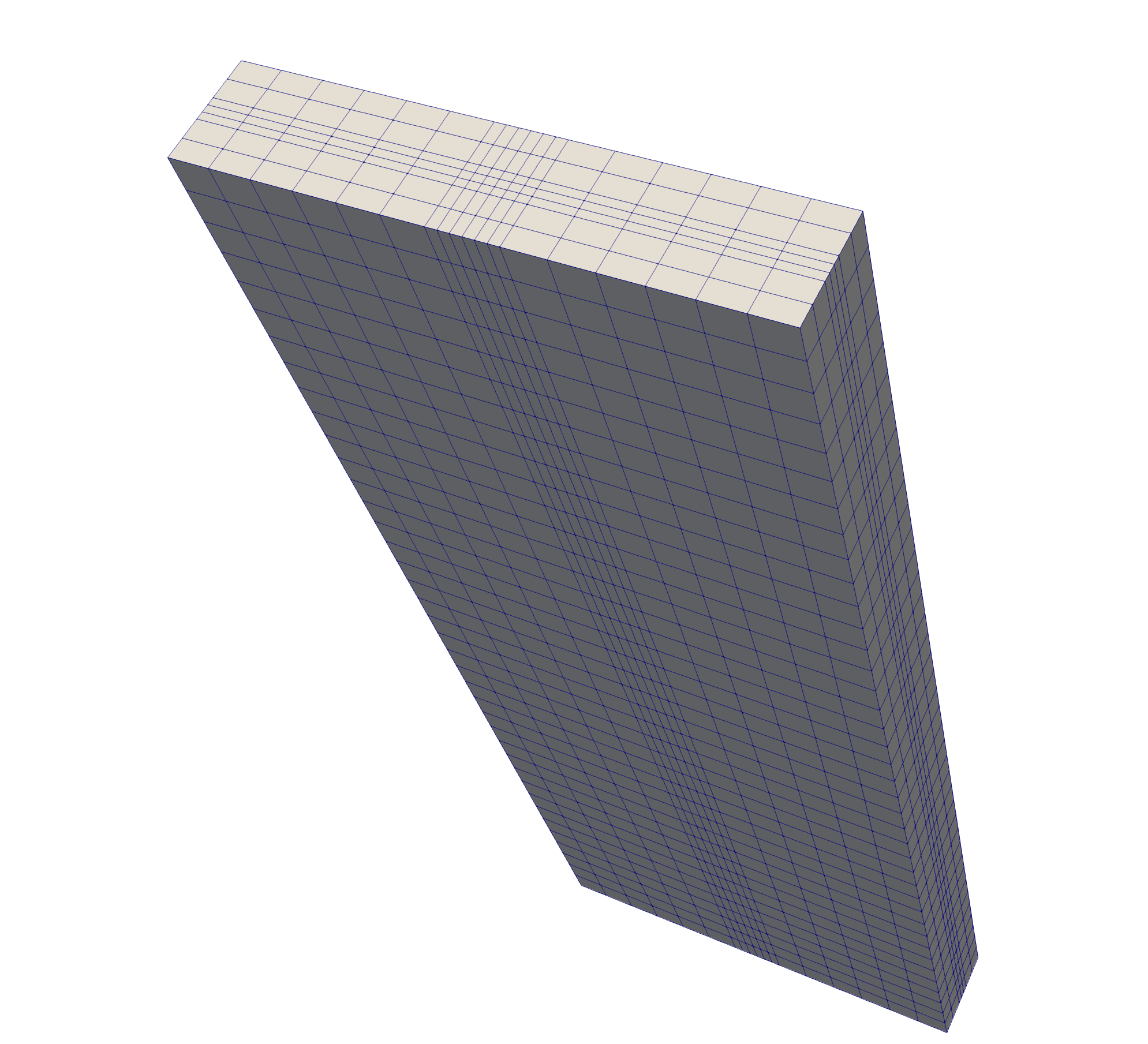}}
\subfloat[Test case A.2]
{\includegraphics
[width=0.3\textwidth]{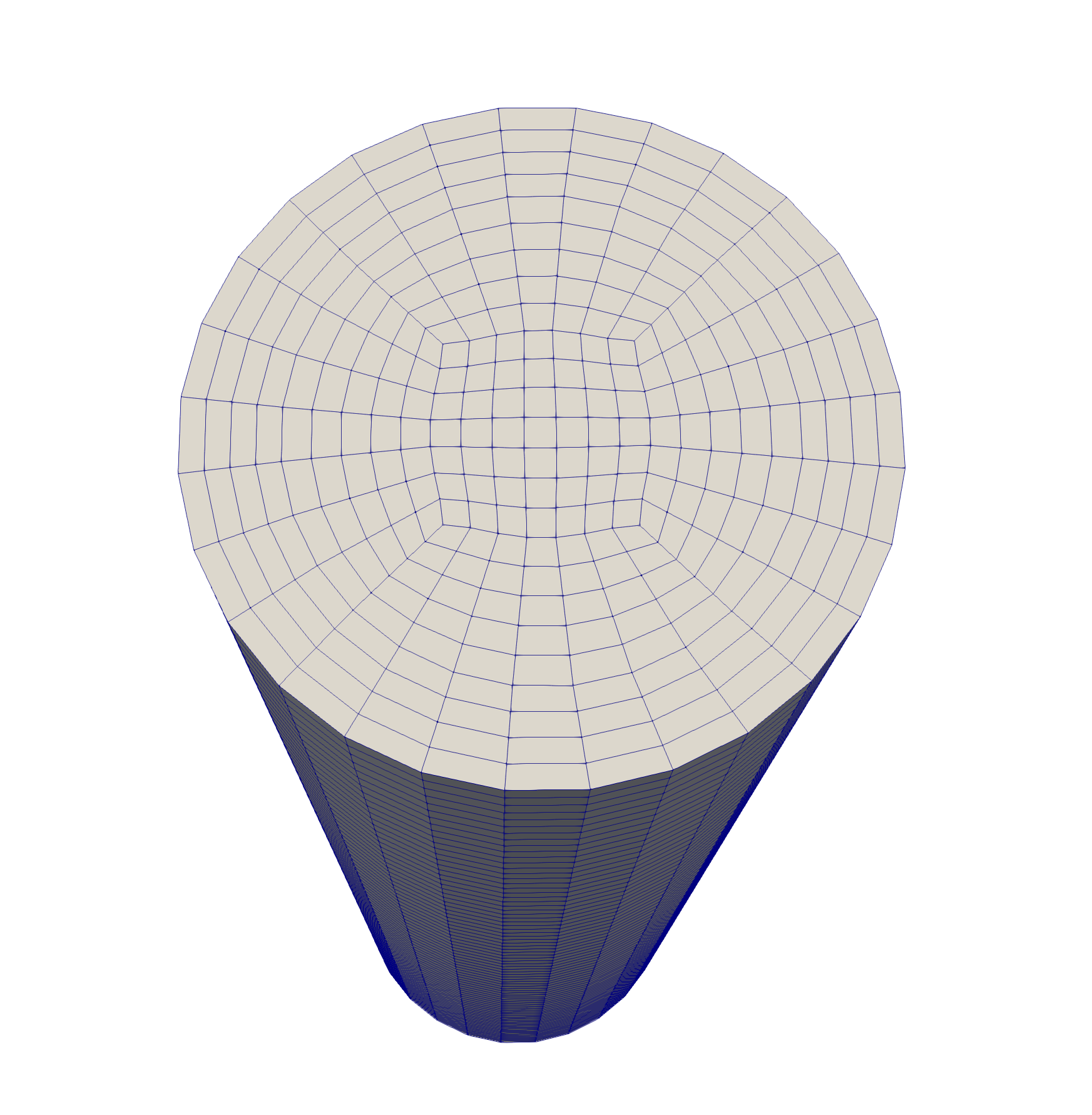}}
\subfloat[Test case A.3]
{\includegraphics
[width=0.3\textwidth]{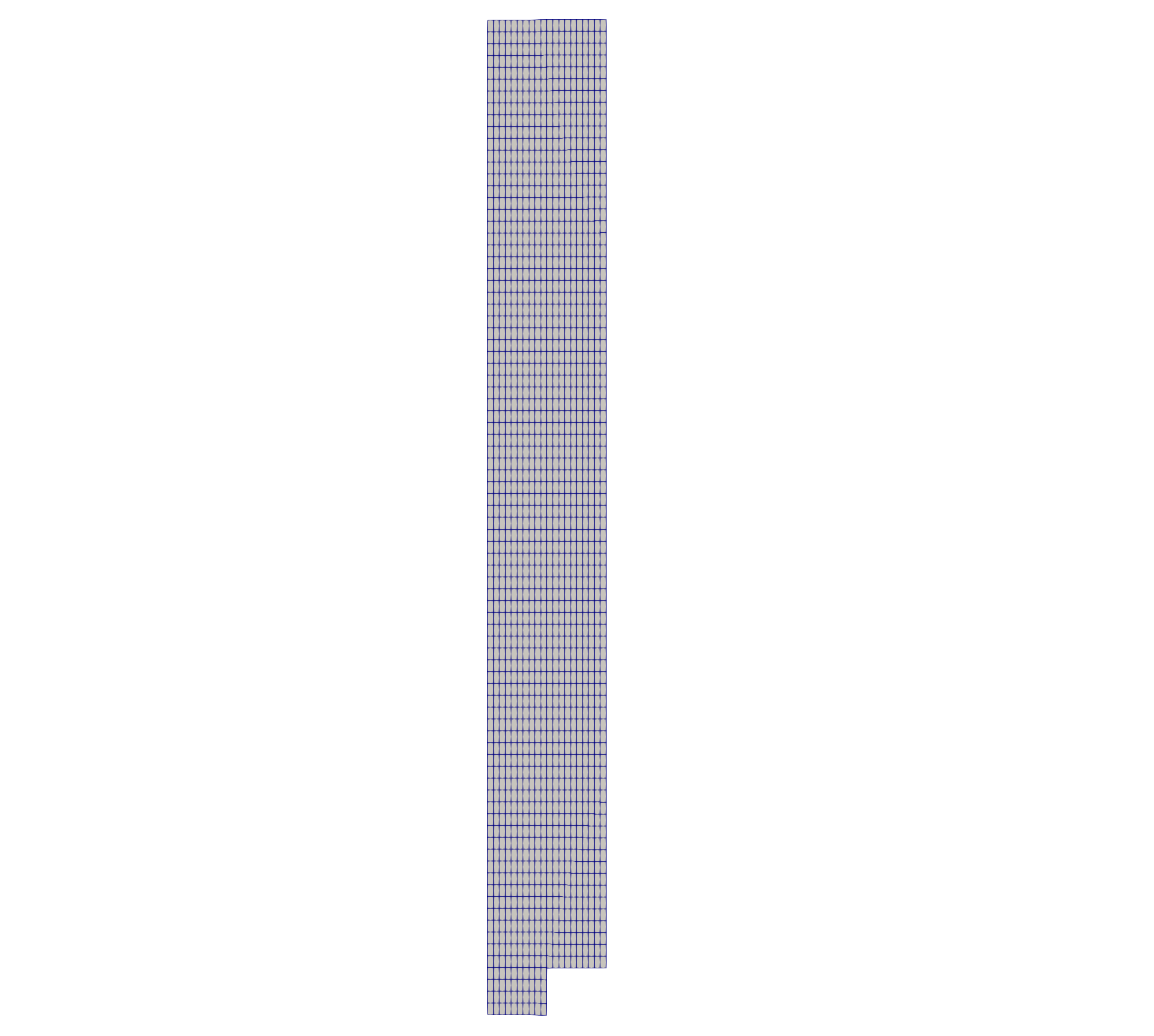}}

\subfloat[Test case B.1]
{\includegraphics
[width=0.3\textwidth]{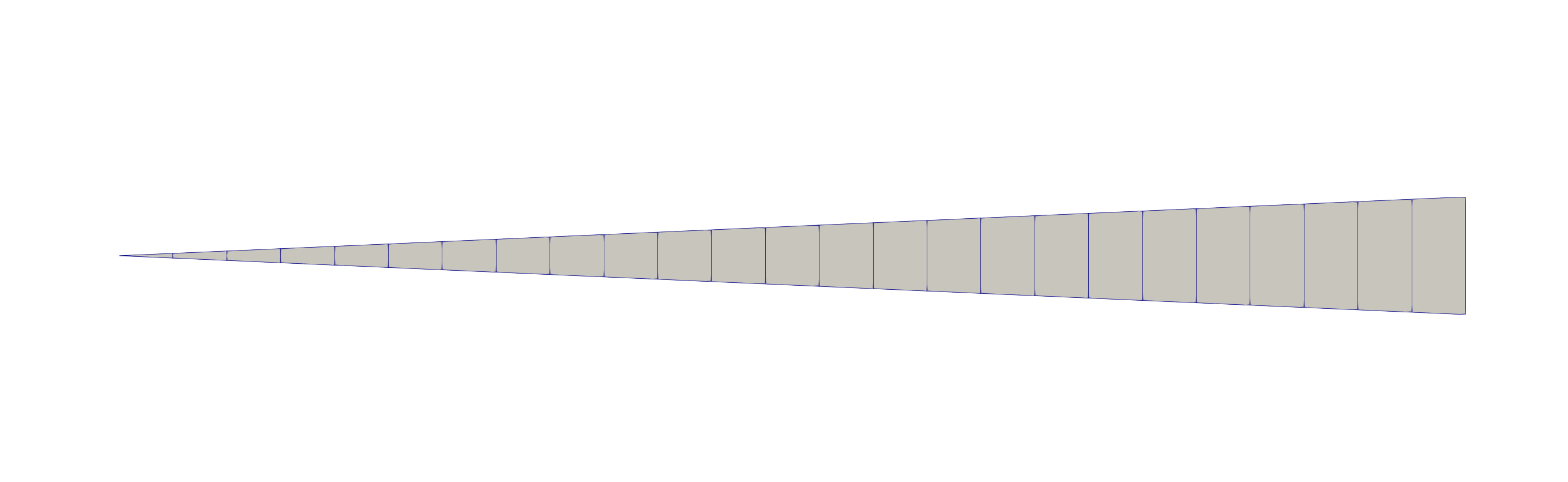}}
\subfloat[Test case B.2]
{\includegraphics
[width=0.3\textwidth]{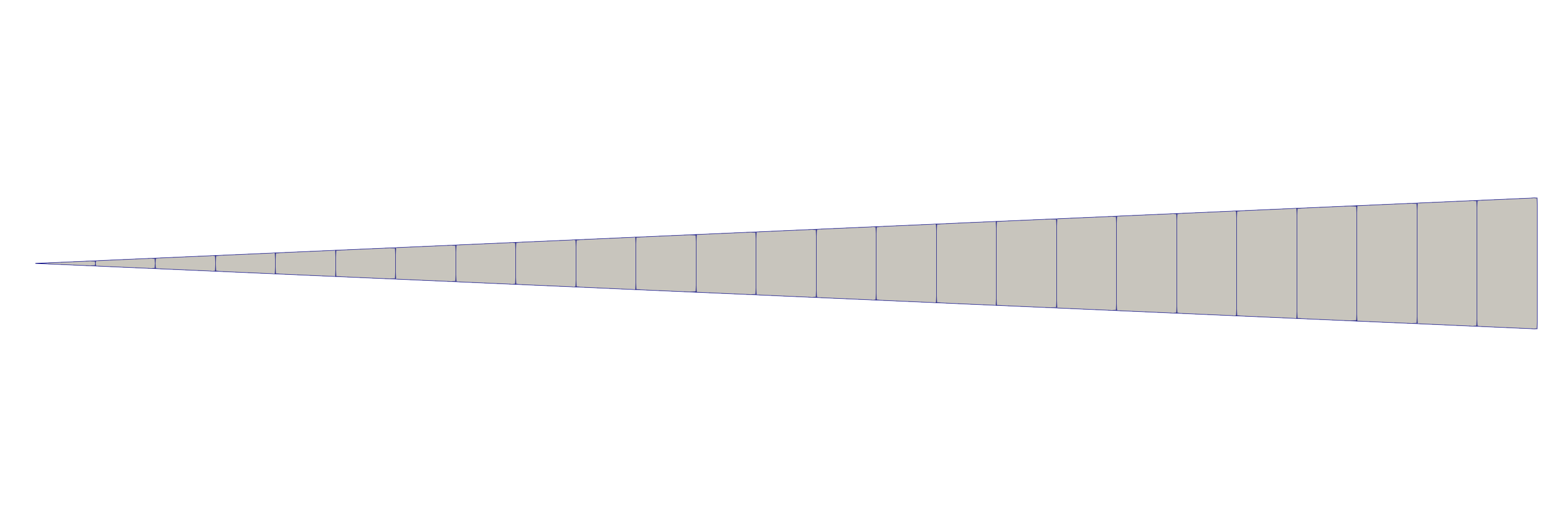}}
\subfloat[Test case B.3]
{\includegraphics
[width=0.3\textwidth]{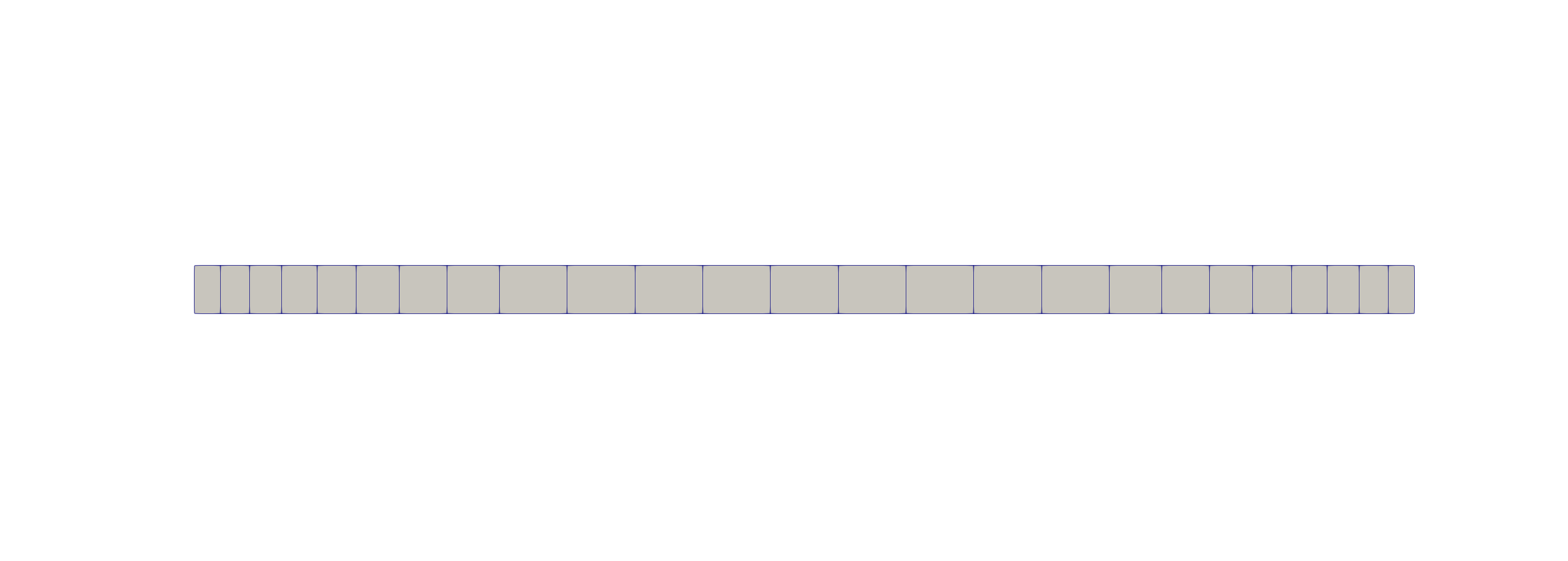}}

\subfloat[Test case B.4]
{\includegraphics
[width=0.3\textwidth]{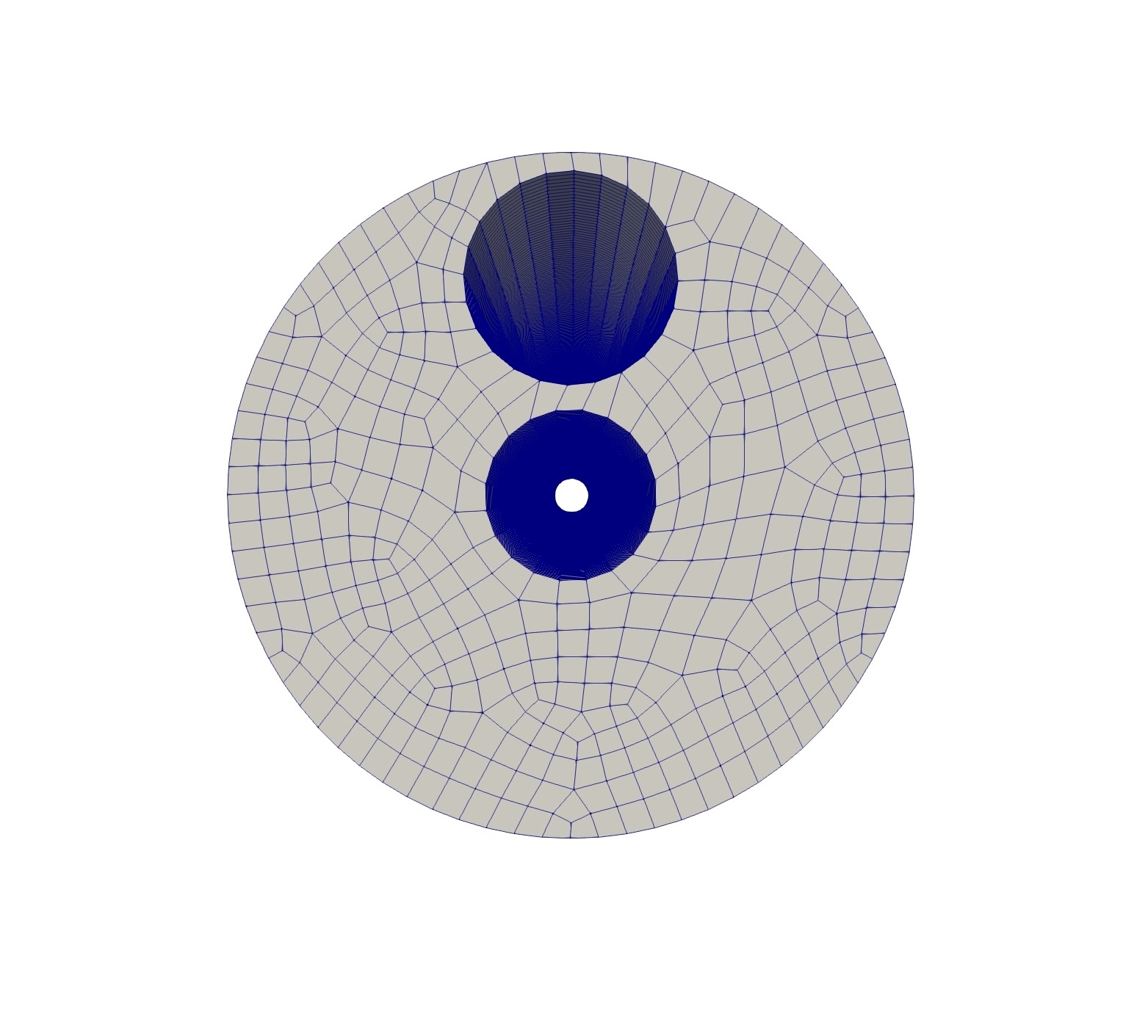}}
\subfloat[Test case C.1]
{\includegraphics
[width=0.3\textwidth]{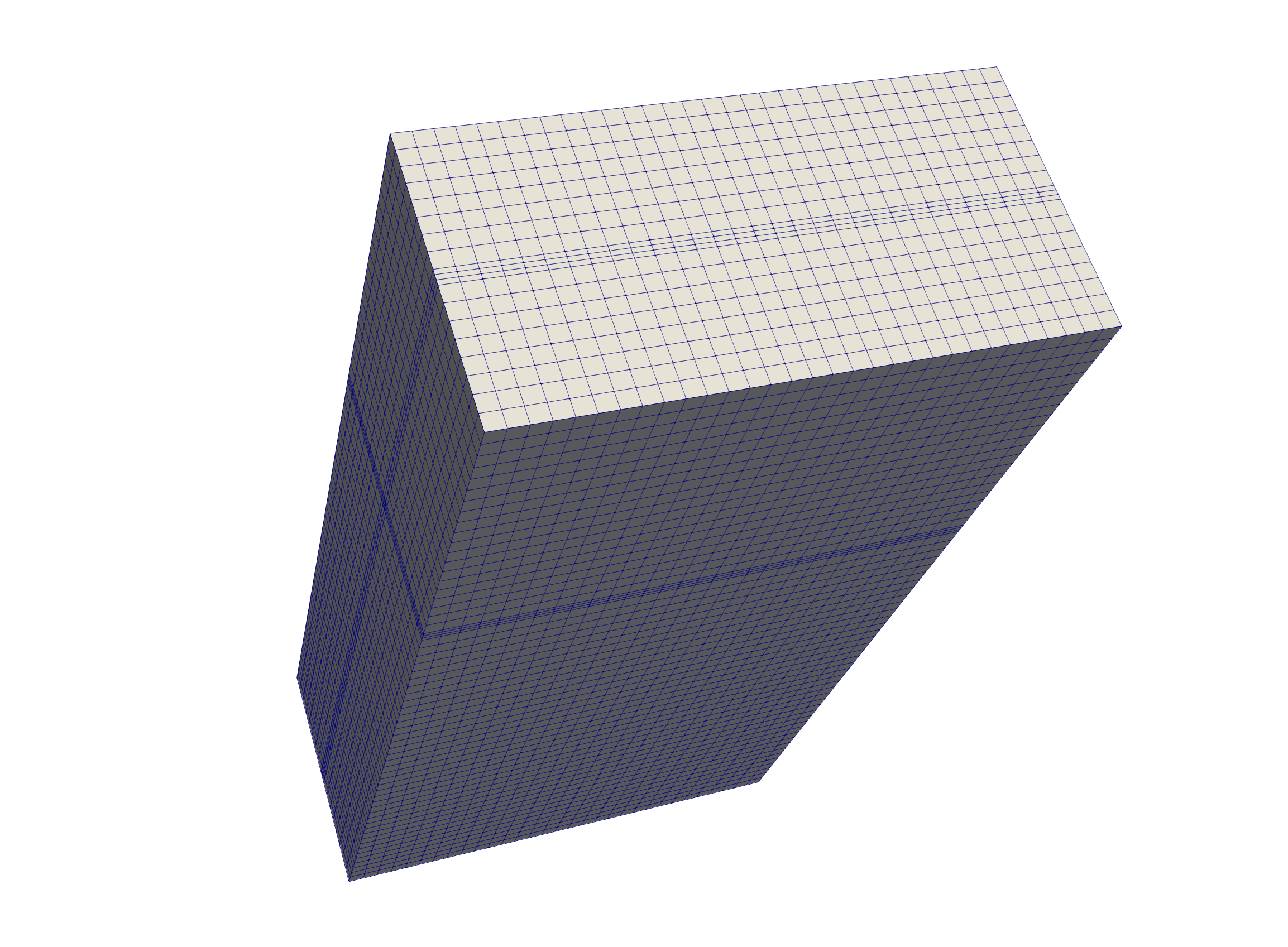}}
\subfloat[Test case C.2]
{\includegraphics
[width=0.3\textwidth]{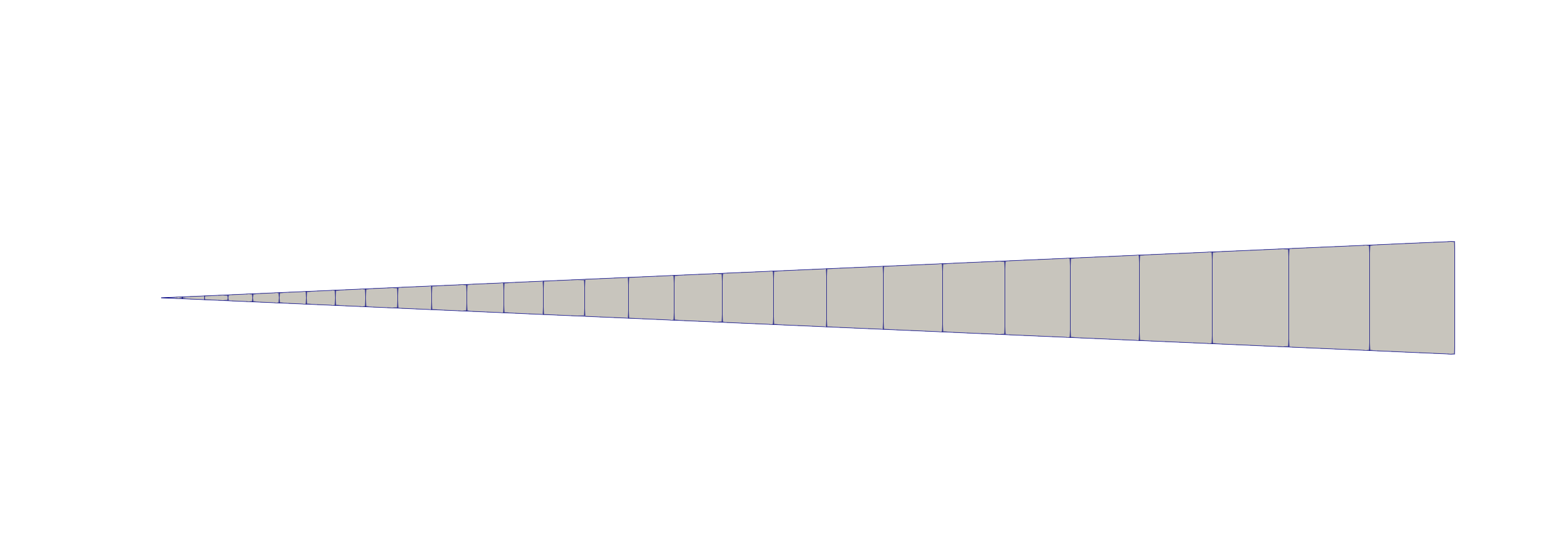}}

\end{center}
\caption{Computational grids employed for the test cases. } 
\label{mesh}
\end{figure}

The computational grids employed in these test cases are shown in Fig. \ref{mesh}. 3D hexahedral cells were generated for test case A.1, A.2 and C.1. 2.5D axi-symmetric wedge cells were employed for test case A.3, B.1, B.2 and C.2. 2D hexahedral cells were employed for test case B.3. Non-regular gambit-paving meshes was employed for test case B.4 due to the existence of the non-regular geometries.  Our grid independence investigation, though not shown herein for brevity, suggested that the predicted results are not sensitive to the grid resolution. The base setup for these test cases are listed in Table \ref{scheme} and \ref{conver}. In our preliminary investigations such numerical setup compromises between stability and accuracy. Therefore, the remainder of the simulations in this study will utilize this base setup. The main boundary conditions are listed in Table \ref{boud}. Since large amount test cases were employed in this work, it is not possible to document all the settings for brevity. Readers are referred to the source code of the test cases for further details. 

All the test cases were simulated by three basic settings. The first one only consists of the drag force. The second one consists of the drag force and turbulence dispersion force. The third one consists of the drag force, turbulence dispersion force, lift force and wall force. Our aim is to verify which force closure combination is able to provide the most universal settings.

\begin{table}[H]
\centering
\begin{tabular}{|l|l|}
\hline
Term & Configuration  \\ \hline
$\p / \p t$ & Euler implicit \\ \hline
$\nabla\psi$ & Gauss linear \\ \hline
$\nabla p$ & Gauss linear \\ \hline
$\nabla\cdot(\psi\bfU\bfU)$ & Gauss limitedLinearV 1;  \\ \hline
$\nabla\cdot(\bfU\psi)$ & Gauss limitedLinear 1 \\ \hline
$\nabla\cdot\tau$ & Gauss linear \\ \hline
$\nabla^2\psi$ & Gauss linear uncorrected \\ \hline
$\nabla^{\perp }\psi$ & uncorrected \\ \hline
$\psi_f$ & linear \\ \hline
\end{tabular}
\caption{Numerical configurations used in the test cases: $\psi$ denotes a generic variable. $(...)_f$ is the face interpolation operator. $\nabla^{\perp }$ is the surface-normal gradient. The number ``1'' indicates the compliance of the scheme with the definition of TVD scheme. A value of 1 indicates full TVD compliance. }
\label{scheme}
\end{table}

\begin{table}[H]
\centering
\begin{tabular}{|c|c|c|c|c|}
\hline
              & Solver    & Preconditioner & Rel. tol. & Final tol.  \\ \hline
$p$           & PCG       & DIC            & 0.01     & 1e-7              \\ \hline
$k$           & PBiCGStab & DILU           & -         & 1e-7           \\ \hline
$\varepsilon$ & PBiCGStab & DILU           & -         & 1e-7           \\ \hline
\end{tabular}
\caption{Solvers and related settings used in the test cases.}
\label{conver}
\end{table}

\begin{table}[H]
\center
\begin{tabular}{|l|l|}
\hline
Test case                      & Operating conditions                                                                                                                                                  \\ \hline
A.1                            & \begin{tabular}[c]{@{}l@{}}Superficial gas velocity: 0.0024 m/s\\ Inlet liquid velocity: 0 m/s\\ Bubble diameter: 0.00505 m\end{tabular}                                    \\ \hline
A.2                            & \begin{tabular}[c]{@{}l@{}}Superficial gas velocity: 0.04 m/s\\ Inlet liquid velocity: 0 m/s\\ Bubble diameter: 0.004 m\end{tabular}                                        \\ \hline
A.3                            & \begin{tabular}[c]{@{}l@{}}Inlet gas velocity: 1.87 m/s\\ Inlet liquid velocity: 1.57 m/s\\ Bubble diameter: 0.002 m\end{tabular}                                      \\ \hline
B.1                            & \begin{tabular}[c]{@{}l@{}}Superficial gas velocity: 0.0115 m/s\\ Superficial liquid velocity: 1.0167 m/s\\ Bubble diameter: 0.0048 m\end{tabular}                                \\ \hline
B.2                            & \begin{tabular}[c]{@{}l@{}}Superficial gas velocity: 0.0151 m/s\\ Superficial liquid velocity: 1.017 m/s\\ Bubble diameter: 0.0046 m\end{tabular}                      \\ \hline
B.3                            & \begin{tabular}[c]{@{}l@{}}Superficial gas velocity: 0.005 m/s\\ Superficial liquid velocity: 0.43 m/s\\ Bubble diameter: 0.006 m\end{tabular}                          \\ \hline
B.4                            & \begin{tabular}[c]{@{}l@{}}Inlet gas velocity: 0.0087 m/s\\ Inlet liquid velocity: 0 m/s\\ Bubble diameter: 0.0042 m\end{tabular}                                     \\ \hline
C.1                            & \begin{tabular}[c]{@{}l@{}}Inlet gas velocity: 4 cm$^3$/s\\ Inlet liquid velocity: 5 l/s\\ Bubble diameter: 0.005 m\end{tabular}                          \\ \hline
C.2                            & \begin{tabular}[c]{@{}l@{}}Inlet gas velocity: 50 ml/s\\ Inlet liquid velocity: 0 m/s\\ Bubble diameter: 0.006 m\end{tabular}                                         \\ \hline
\multicolumn{2}{|l|}{\begin{tabular}[c]{@{}l@{}}Gas velocity at walls: slip. Liquid velocity at walls: no-slip. \\ $k$ and $\varepsilon$ at walls: wall function. Outlet: zero-gradient.\end{tabular}} \\ \hline
\end{tabular}
\caption{Main boundary conditions adopted in the simulations. } 
\label{boud}
\end{table}

\section{Results and discussions}

\subsection{Test case A.1 - A.3}
In this section, the numerical predicted results for test case A.1 to A.3 were investigated. The available experimental data and features are listed in Table \ref{AA}. These test cases are common in chemical and bio-processing engineering. In test case A.1, the gas phase was injected into the column at a relatively small superficial velocity. A ``cooling tower'' flow regime was formed due to the existence of periodic large vortex. In test case A.2, the gas phase was injected into the cylinder column at high superficial velocities. The flow fields were highly transient and full of vortexes. In test case A.3, the gas phase was injected into a sudden enlargement pipe and a steady-state stagnant vortex was formed at the pipe bottom. In experiments, the turbulence kinetic energy was monitored, which can be used to validate the turbulence model employed in the E-E method. 
\begin{table}[H]
\center
\begin{tabular}{|c|c|c|c|}
\hline
Test case & Exp. data                                                                  & Features                                                                       & Pseudo-steady-state \\ \hline
A.1       & \begin{tabular}[c]{@{}c@{}}Plume oscillating period\\ Gas holdup\end{tabular} & Periodic flow field                                                                      & No                 \\ \hline
A.2       & Phase frac. distri.                                                           & High phase fraction                                                                    & No                \\ \hline
A.3       & \begin{tabular}[c]{@{}c@{}}Upward liquid vel.\\ Turb. kinetic energy\end{tabular} & Stagnant vortex & Yes                 \\ \hline
\end{tabular}
\caption{Available experimental data and features of test case A.1 to A.3.}
\label{AA}
\end{table}

\begin{figure}[H]
\begin{center}
\includegraphics
[width=0.32\textwidth]{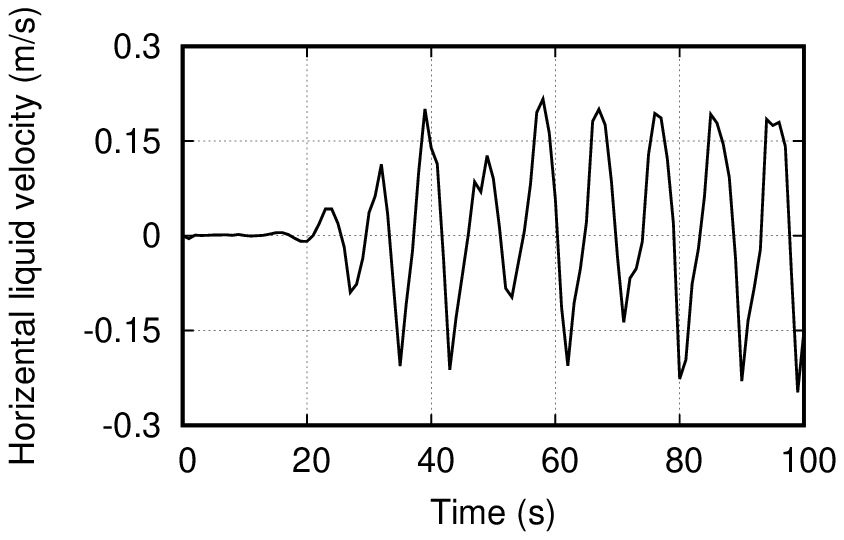}
\includegraphics
[width=0.32\textwidth]{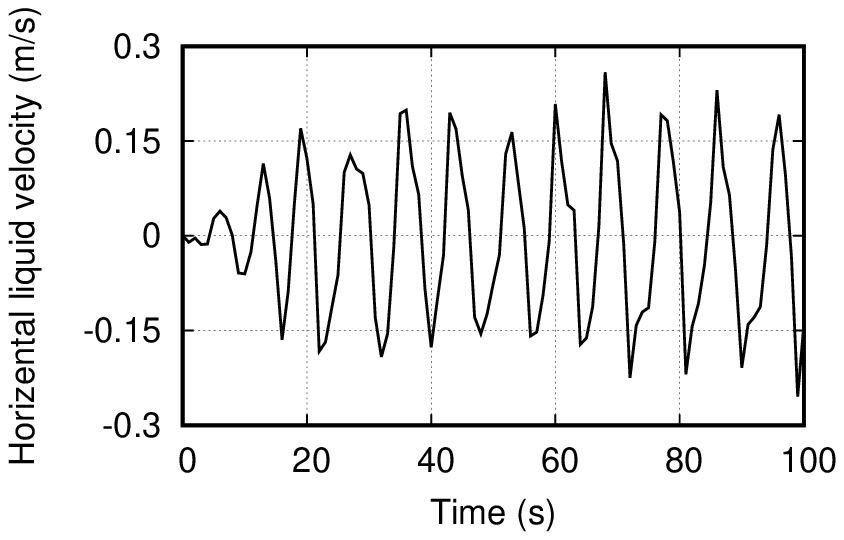}
\includegraphics
[width=0.32\textwidth]{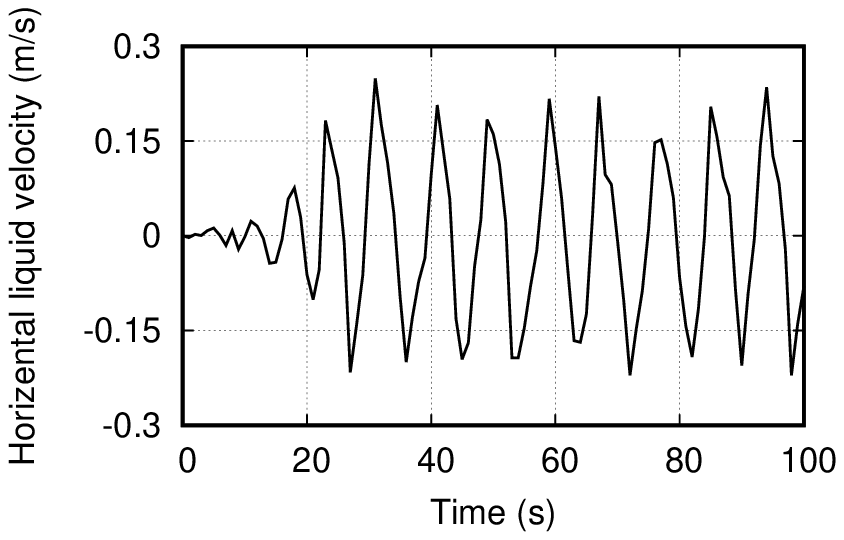}
\end{center}
\caption{Horizontal liquid velocity predicted for case A.1. Left: drag force. Middle: drag force and turbulent dispersion force. Right: drag force, turbulent dispersion force, lift force and wall force. $\bfU_g=0.0024$ m/s. Location: $x$ = 0.1 m, $y$ = 0.225 m, $z$ = 0.02 m.} 
\label{caseA1}
\end{figure}

Fig. \ref{caseA1} show the horizontal liquid velocity predicted by different force closure combination for test case A.1. It can be seen that all the force closure combination can predict the periodically ``Gulf-stream'' phenomenon, which proves that the periodically vortex in the bubble column is not sensitive to the force closure. Table \ref{caseA13} shows the predicted gas holdup and the plume oscillation period (POP). Compared with the experimental data, the POP predicted by all these force closure combination was under-estimated. Results can be improved by adjusting the turbulence model and inclusion of the bubble induced turbulence as was shown in our previous work \cite{li2017simulation}. On the other hand, all these three force closure combination is able to predict good results of the gas holdup with errors smaller than 11\%. When the turbulence dispersion force and wall forces were included, the prediction of the gas holdup was slightly improved. However, the addition of wall forces and turbulence dispersion forces cannot improve the results of POP. It can be explained by the fact that the phase shear rate in this test case is too small to present differences, and the occurence of POP (or the pediodically vortex) is mainly because of the drag.

\begin{table}[H]
\begin{center}
\begin{tabular}{|l|l|l|l|l|l|l|}
\hline
          & Drag force  & Drag \& dispersion forces     & \begin{tabular}[c]{@{}c@{}} Drag \& dispersion \\  \& lift and wall forces \end{tabular} & Exp.   \\ \hline
Gas holdup & 0.00613      & 0.0065 & 0.0075 & 0.0069 \\ \hline
\begin{tabular}[c]{@{}c@{}} Gas holdup\\ rel. error\end{tabular}  & -11\%     &-6\%  & +8\%                                                               & - \\ \hline
POP & 9.37   & 8.87  & 8.98 & 11.38 \\ \hline
POP rel. error & -17\%   & -22\% & -21\% & - \\ \hline
\end{tabular}
\end{center}
\caption{Comparison of the gas holdup and POP predicted by different force closure combination with experimental data for test case A.1.} 
\label{caseA13}
\end{table}

\begin{figure}[H]
\begin{center}
\includegraphics
[width=0.48\textwidth]{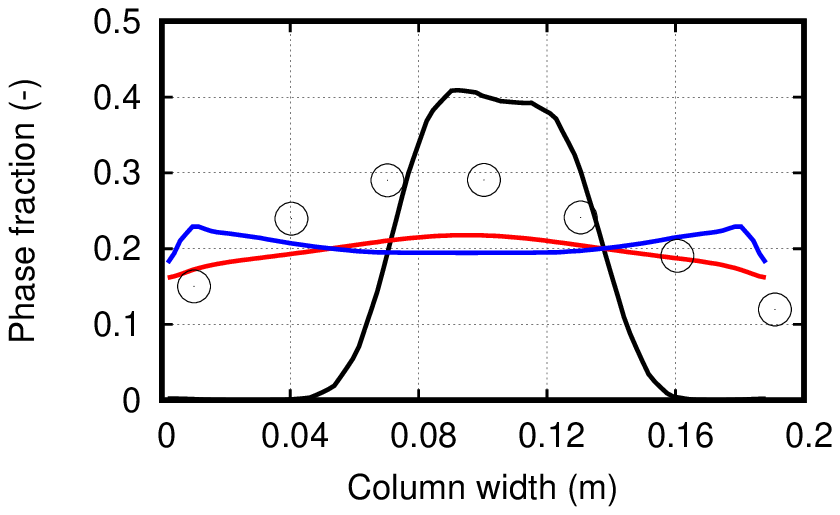}
\includegraphics
[width=0.48\textwidth]{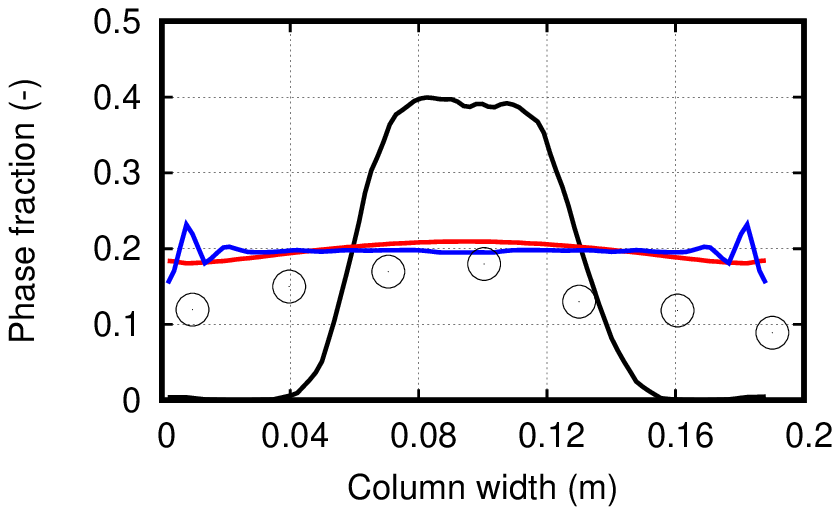}
\end{center}
\caption{Comparison of the phase fraction (lines) predicted by different force closure combination with experimental data (circles). $\bfU_g=0.04$ m/s. Black line: drag force. Red line: drag force and turbulent dispersion force. Blue line: drag force, turbulent dispersion force, lift and wall forces. } 
\label{caseA2}
\end{figure}

Fig. \ref{caseA2} shows the averaged phase fraction predicted by different force closure combinations for test case A.2. It can be seen that the phase fraction is highly dependent on the employed force models and the difference is quite large. If only the drag force was included, the predicted phase fraction was accumulated at the center of the column since no lateral forces and diffusion force were used. If the turbulence dispersion force was included, the bubbles at the center of the column were diffused along the gradient of the phase fraction and the bubble accumulating problem was prevented.  Although the predicted phase fraction was flattened, it was much better than the predicted results when only the drag force was included. Moreover, the flattened phase fraction can be handled by using a smaller the turbulence dispersion force coefficients. The addition of the wall forces cannot improve the results. Instead, the wall forces predict a peak in the vicinity of the wall, which was not observed in the experiments.

\begin{figure}[H]
\begin{center}
\includegraphics
[width=0.32\textwidth]{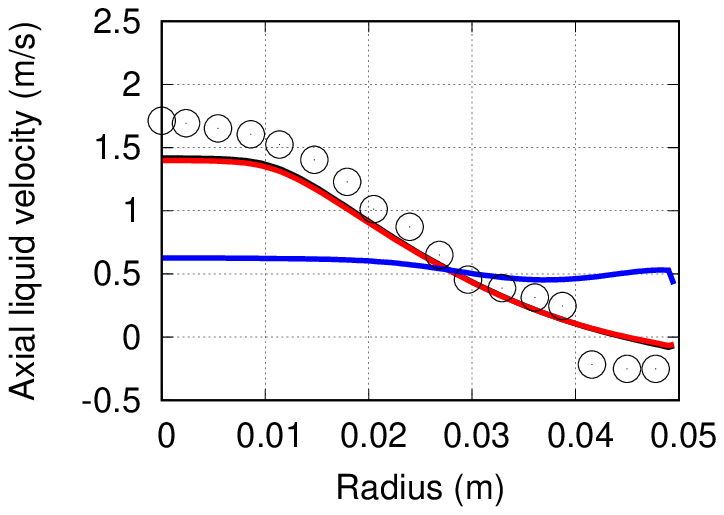}
\includegraphics
[width=0.32\textwidth]{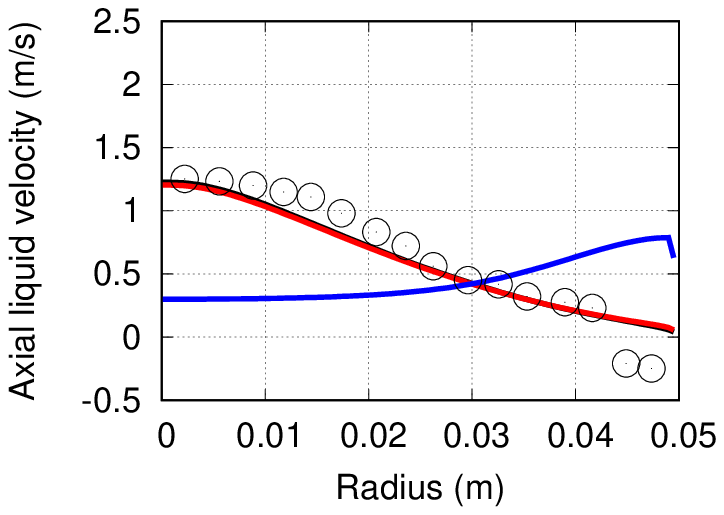}
\includegraphics
[width=0.32\textwidth]{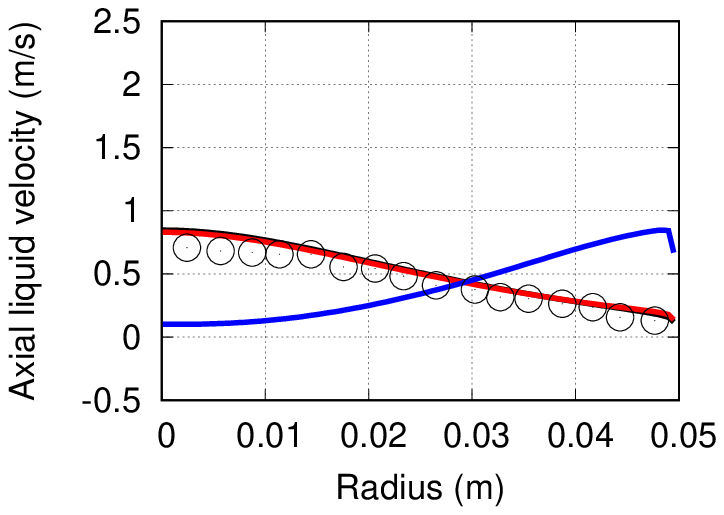}
\includegraphics
[width=0.32\textwidth]{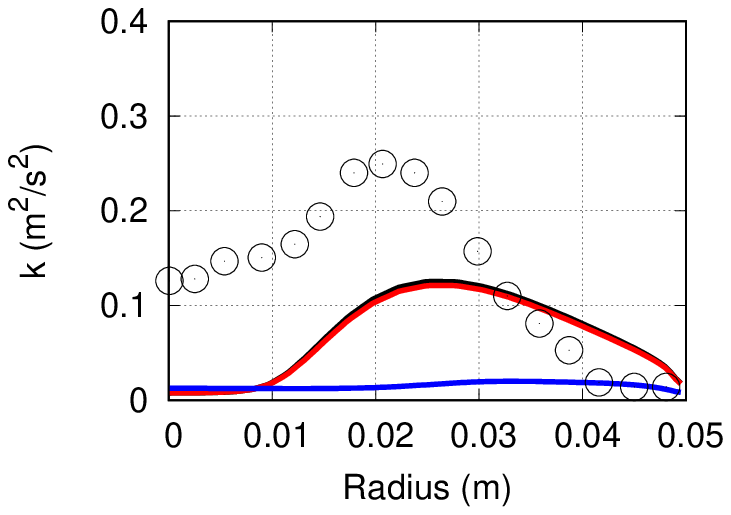}
\includegraphics
[width=0.32\textwidth]{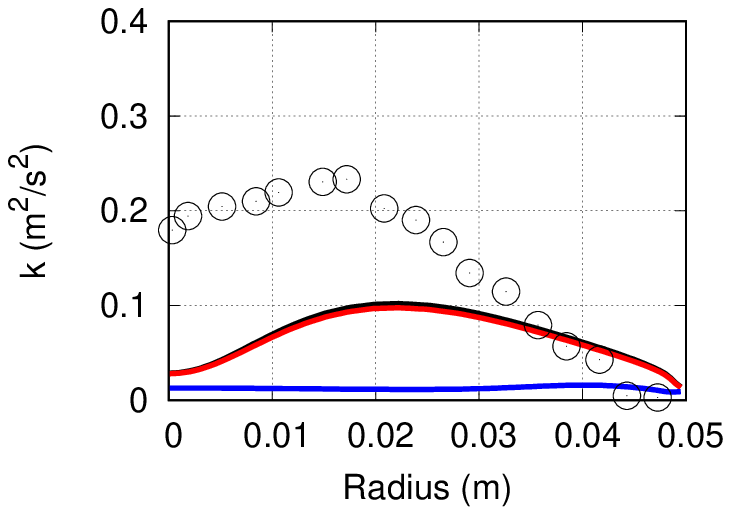}
\includegraphics
[width=0.32\textwidth]{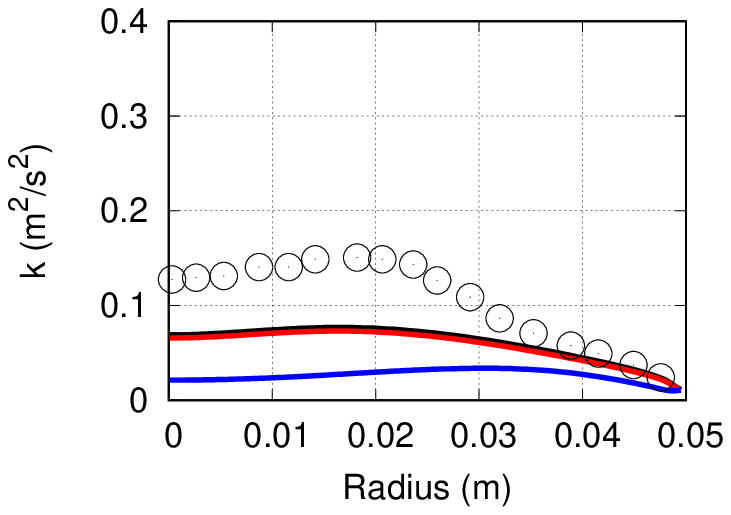}
\end{center}
\caption{Predicted profiles of the mean axial liquid velocity (top) and the turbulent kinetic energy (bottom) compared with experimental data (circles) at different height for case A.3. Left: $y=0.15$ m, middle:  $y=0.25$ m,  right $y=0.32$ m. Red line: drag force. Black line: drag force and turbulent dispersion force. Blue line: drag force, turbulent dispersion force, lift and wall forces. } 
\label{caseA3U}
\end{figure}

Fig. \ref{caseA3U} shows the predicted mean axial liquid velocities and the turbulent kinetic energy for test case A.3. It can be seen that the prediction of the axial liquid velocity agree well with experiments when the drag force and turbulence dispersion force were included. However, including the wall forces cannot predict reasonable results. Meanwhile, accurate prediction of the turbulent kinetic energy was proven to be quite difficult. Although the combination of the drag force and turbulence dispersion force improves the results compared with that predicted by addition of the wall forces, the predicted turbulent kinetic energy was under-estimated. Further research is
needed to quantify these errors and possibly employ a bubble induced turbulence model to correct the deficiencies \cite{vaidheeswaran2017bubble}.

\subsection{Test case B.1 - B.4}
In this section, the numerical predicted results for test case B.1 to B.4 were investigated. The available experimental data and features are listed in Table \ref{BB}. These test cases are common in nuclear engineering. In test case B.1 to B.3, the aspect ratios of the geometry are quite large, which implies that there is enough time for the bubbles and the phase fraction to develop along the vertical pipe direction. Pseudo-steady-state can also be reached due to small gas phase fraction and low gas holdup. In test case B.4, non-regular internal pipes are placed in the computational domain and the aspect ratio is relatively small. Meanwhile, it was found in the experiments that the flow field in the pipe is transient due to the existence of the non-regular internal pipes. These test cases can be distinguished by different features. In our preliminary investigations as mentioned previously, we found that the non-drag forces are crucial to obtain reasonable radial phase fraction distribution, especially for the pseudo-steady-state test cases (e.g., case B.1 to B.3). Therefore, in the following, we will turn our attention to the non-drag forces.

\begin{table}[H]
\center
\begin{tabular}{|c|c|c|c|}
\hline
Test case & Exp. data                                                                  & Features                                                                       & Pseudo-steady-state \\ \hline
B.1       & \begin{tabular}[c]{@{}c@{}}Phase frac. distri.\\ Upward gas vel.\end{tabular} & Wall peak                                                                      & Yes                 \\ \hline
B.2       & Phase frac. distri.                                                           & Double peak                                                                    & Yes                 \\ \hline
B.3       & Phase frac. distri.                                                           & Central peak & Yes                 \\ \hline
B.4       & Global gas holdup                                                          & Non-regular components                                                              & No \\ \hline
\end{tabular}
\caption{Available experimental data and features of test case B.1 to B.4.}
\label{BB}
\end{table}

\begin{figure}[H]
\begin{center}
\includegraphics
[width=0.48\textwidth]{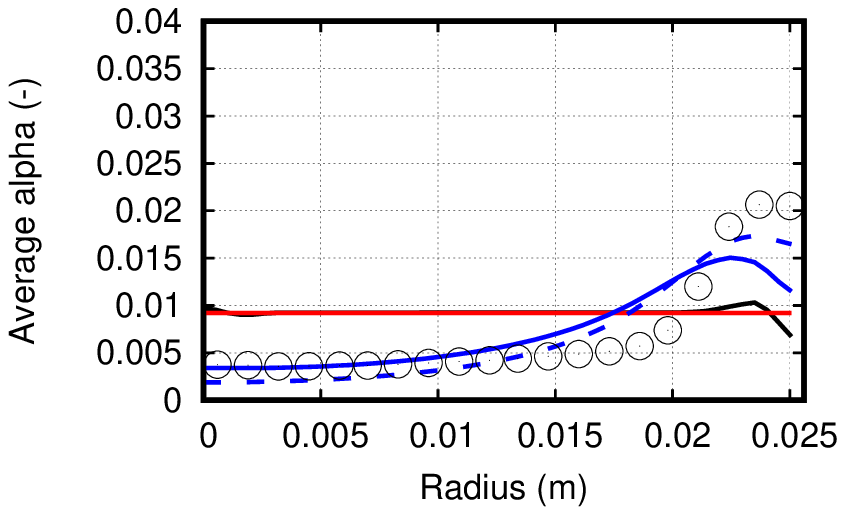}
\includegraphics
[width=0.48\textwidth]{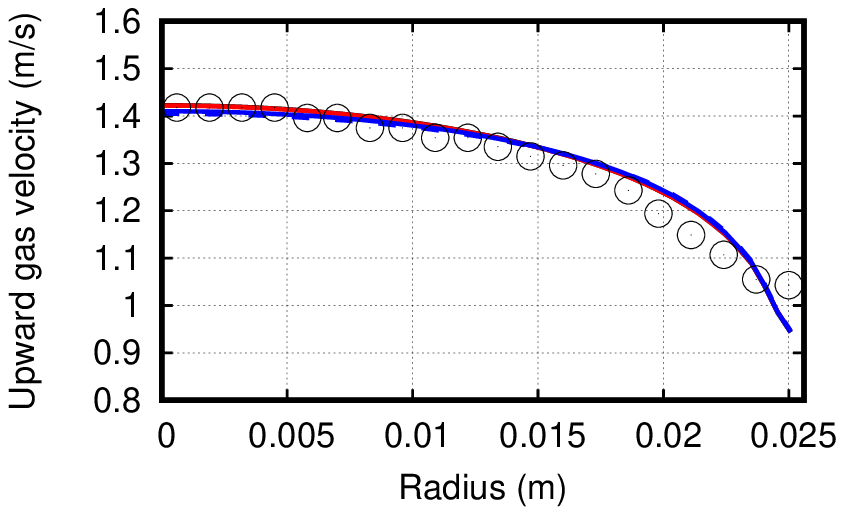}
\end{center}
\caption{Comparison of the upward gas velocity (left, lines) and phase fraction (right, lines) with experimental data (circles) for case B.1. Sample location: height of 3.03 m above the inlet. Red line: drag force. Black line: drag force and turbulent dispersion force. Blue solid line: drag force, turbulent dispersion force, lift and wall forces, $C_2=0.05$. Blue dashed line: $C_2=0.03$. } 
\label{caseB1}
\end{figure}

Fig. \ref{caseB1} shows the comparison of the upward gas velocity and phase fraction compared with experimental data. It can be seen that the bubble upward velocity is not sensitive to the force closure. These force combination predict similar results. Accurate predictions of the upward gas velocity can be also obtained if only the drag force was included. However, reasonable phase fraction can only be predicted when all the forces was included. Meanwhile, we found that the predicted phase fraction was highly dependent on the wall lubrication force model. For the wall lubrication  model developed by Antal et al. \cite{antal1991analysis}, using a smaller $C_2$ improved the results. Otherwise many bubbles are pushed away when a large $C_2$ was used.

\begin{figure}[H]
\begin{center}
\includegraphics
[width=0.6\textwidth]{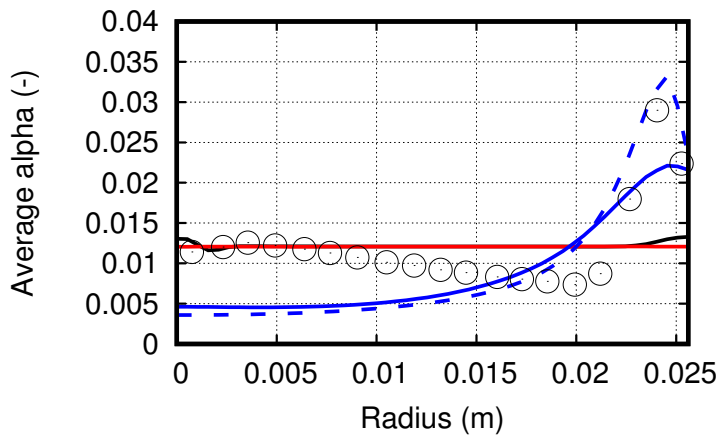}
\end{center}
\caption{Predicted profiles of the phase fraction (lines) compared with experimental data (circles) for case B.2. $\bfU_g=0.0151$ m/s. $\bfU_l=0.534$ m/s.  Location: $L/D=59$. Red line: drag force. Black line: drag force and turbulent dispersion force. Blue solid line: drag force, turbulent dispersion force, lift and wall forces, $C_2=0.05$. Blue dashed line: drag force, turbulent dispersion force, lift and wall forces, $C_2=0.03$. } 
\label{caseB2}
\end{figure}

\begin{figure}[H]
\begin{center}
\includegraphics
[width=0.6\textwidth]{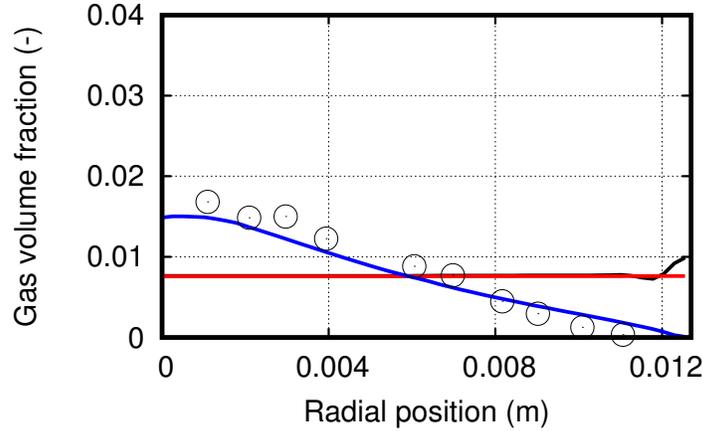}
\end{center}
\caption{Predicted profiles of the phase fraction (lines) compared with experimental data (circles) for case B.3. Location: $z=1.143$ m. Red line: drag force. Black line: drag force and turbulent dispersion force. Blue solid line: drag force, turbulent dispersion force, lift and wall forces. } 
\label{caseB3}
\end{figure}

Fig. \ref{caseB2} shows the radial profiles of the predicted phase fraction for test case B.2. The difference between test case B.1 and B.2 is that a double peak was observed in the latter one due to the existence of bubbles with different diameters. In the mono-disperse plot, the phase fraction of the small bubbles ($d<5.6$ mm) was reported. All these small bubbles tend to move towards the wall and the wall peak was observed. As mentioned previously, reasonable phase fraction can only be predicted when all the forces was included. By using a smaller $C_2$, the wall peak of the phase fraction can be improved. However, none of these models predicted the central peak. It comes from the drawback of the E-E method. As a mono-disperse mathematical model, it can only be used for mono-disperse test cases. Improvements can be obtained by using a higher-order moments methods as it was done in our previous works  \cite{li2019comparison,li2019twowaygpbefoam}. Fig. \ref{caseB3} shows the radial profiles of the predicted phase fraction for test case B.3. In this test case, a large bubble diameter was used. Therefore, they move towards the pipe center due to negative lift force coefficients. It can be succeffully predicted by the lift force model developed by Tomiyama et al. \cite{tomiyama2002transverse}. Again, reasonable phase fraction can only be predicted when all the forces was included.

Table \ref{caseB4} reported the predicted gas holdup compared with the experimental data for test case B.4. It can be seen that the predictions were slightly under-estimated. Results may be improved by taking polydispersity into account as was done in the work of Besagni et al.  \cite{besagni2016annular}. However, even the gas holdup was under-estimated, it can be seen that the prediction was improved when all the forces were included. If only the drag force was included, it results in the largest error.

\begin{table}[H]
\center
\begin{tabular}{|l|l|l|l|l|l|}
\hline
           & Drag  & Drag \& Dis. & \begin{tabular}[c]{@{}l@{}}Drag \& Dis.\\ \& Lift \& Wall\end{tabular} & Exp.   \\ \hline
Gas holdup & 0.0226 & 0.02318       & 0.02358                                                                & 0.0287 \\ \hline
\end{tabular}
\caption{Predicted gas holdup compared with the experimental data for case B.4.} 
\label{caseB4}
\end{table}
\subsection{Test case C.1 - C.2}

Test case C.1 and C.2 are common in metallurgical industry. The available experimental data and features are listed in Table \ref{CC}. In test case C.1, the gas phase is injected into the mold by a submerged entrance nozzle (SEN). Large bubbles are lifted toward the meniscus due to the buoyancy force acting on them and removed from the mold, which smaller bubbles are carried deep into the mold. These small bubbles are trapped in the steel and cause pin hold defects. In the field of numerical simulation, this is also called phase segregation which is very difficult to address. In test case C.2, the gas phase was injected to remove the non-metallic inclusions in metallurgical reactor. This test case is important because it is the only one for which the turbulent kinetic energy was reported in the experiments.

\begin{table}[H]
\center
\begin{tabular}{|c|c|c|c|}
\hline
Test case & Exp. data                                                                  & Features                                                                       & Pseudo-steady-state \\ \hline
C.1       & \begin{tabular}[c]{@{}c@{}}Axial liquid vel.\\ Radial liquid vel.\end{tabular} & Non-regular flow                                                                      & Yes                 \\ \hline
C.2       &\begin{tabular}[c]{@{}c@{}}Axial liquid vel.\\ Turbulent kinetic energy\end{tabular}& Central bubble plume           & Yes                 \\ \hline
\end{tabular}
\caption{Available experimental data and features of test case C.1 to C.2.}
\label{CC}
\end{table}

\begin{figure}[H]
\begin{center}
\includegraphics
[width=0.48\textwidth]{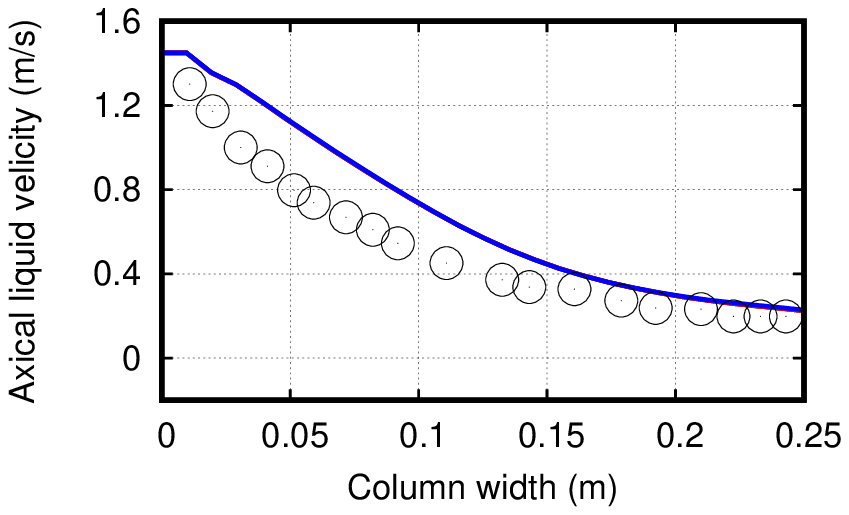}
\includegraphics
[width=0.48\textwidth]{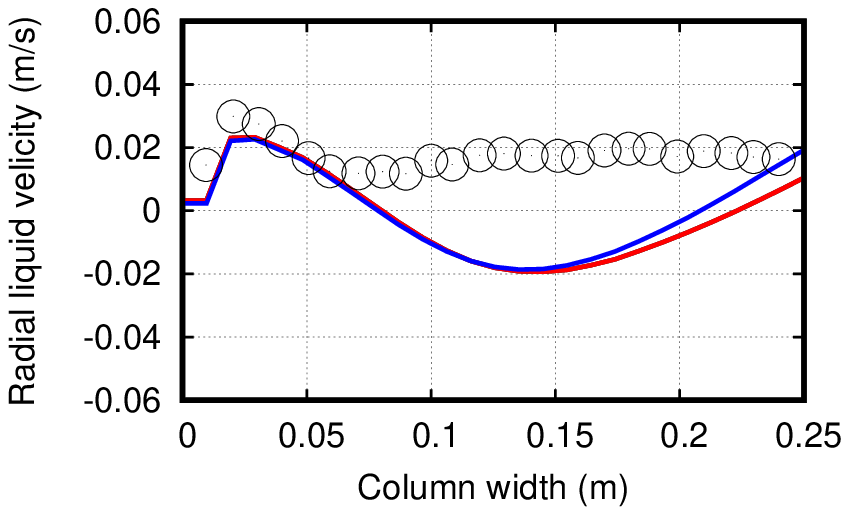}
\end{center}
\caption{Comparison of the axial and racial liquid velocity predicted by different force closure combination with experimental data (points) for case C.1. Red line: drag force. Black line: drag force and turbulent dispersion force. Blue solid line: drag force, turbulent dispersion force, lift and wall forces. Predictions oh the $u$ and $v$ components were made along the center line of the inlet boundary condition. } 
\label{caseC1}
\end{figure}

Fig. \ref{caseC1} shows the predicted axial and radial liquid velocities. It can be seen that all the force closure combination can predict accurate axial liquid velocity compared with experimental data. The predicted axial liquid velocity agree well with the experimental data. On the other hand, none of the models can predict satisfactory radial liquid velocity. Such inconsistency may come from the mono-disperse assumption of the E-E method and can be handled by using a poly-disperse model. The predicted racial liquid velocities at the right tail of the plot were smaller than 0, which implies the liquid is moving toward the bottom of the mold and a large vortex at the right bottom of the mold is formed. Using a larger bubble diameter improves the predictions due to the effect of large buoyancy.  Similar unrealistically predictions for low gas flow rate were also found by Liu and Li \cite{liu2017large}, in which the flow fields were investigated by large eddy simulation. 

\begin{figure}[H]
\begin{center}
\subfloat[]
{\includegraphics
[width=0.45\textwidth]{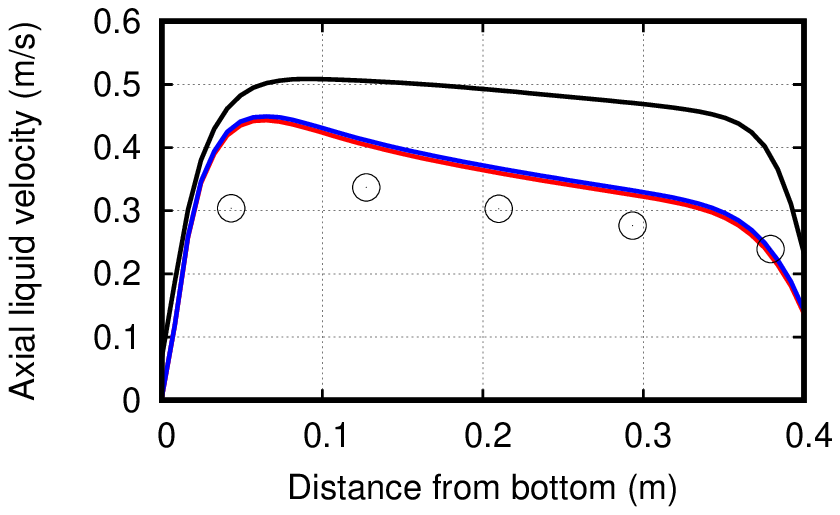}}
\subfloat[]
{\includegraphics
[width=0.45\textwidth]{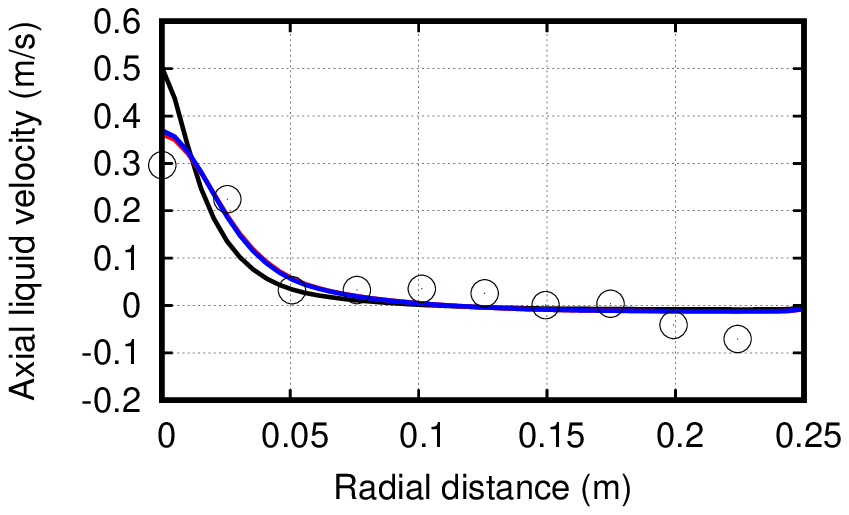}}

\subfloat[]
{\includegraphics
[width=0.45\textwidth]{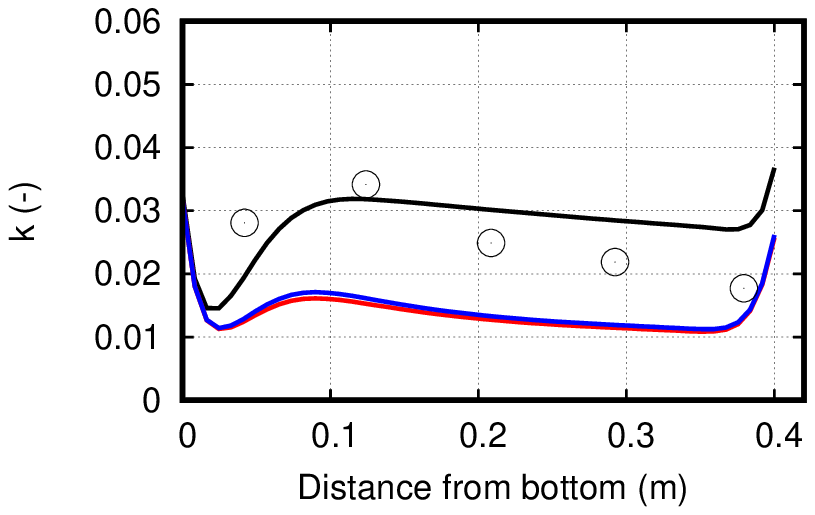}}
\subfloat[]
{\includegraphics
[width=0.45\textwidth]{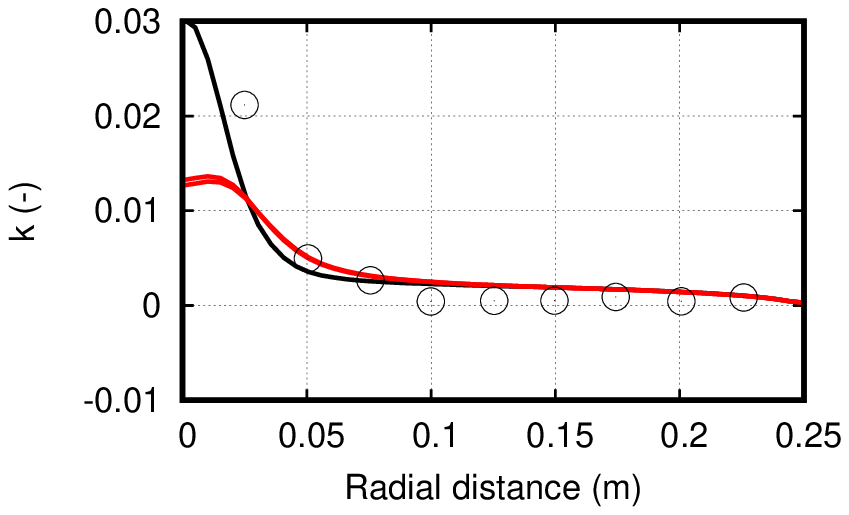}}

\subfloat[]
{\includegraphics
[width=0.45\textwidth]{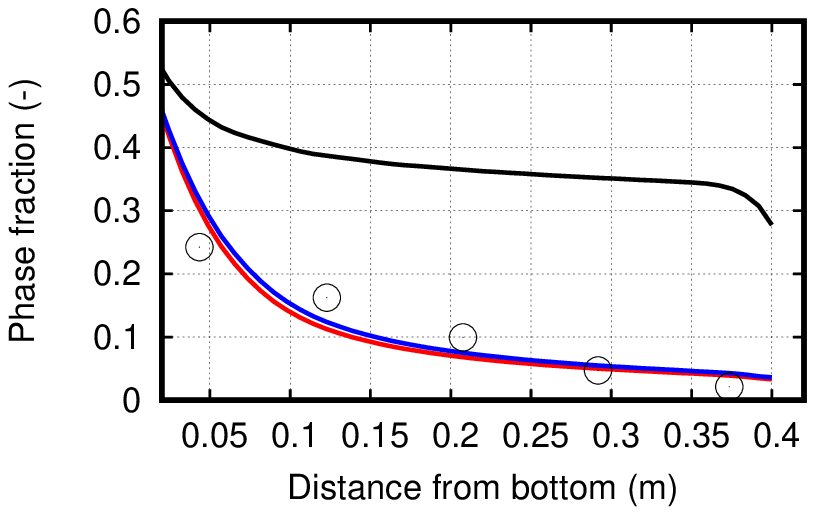}}
\end{center}
\caption{Comparison of the predicted axial liquid velocity and turbulent kinetic energy (lines) and experimental data (points) for case C.2. Location: along the center line of the bubble plume. Red line: drag force. Black line: drag force and turbulent dispersion force. Blue solid line: drag force, turbulent dispersion force, lift and wall forces. } 
\label{caseC2sheng}
\end{figure}

\begin{figure}[H]
\begin{center}
\includegraphics
[width=0.48\textwidth]{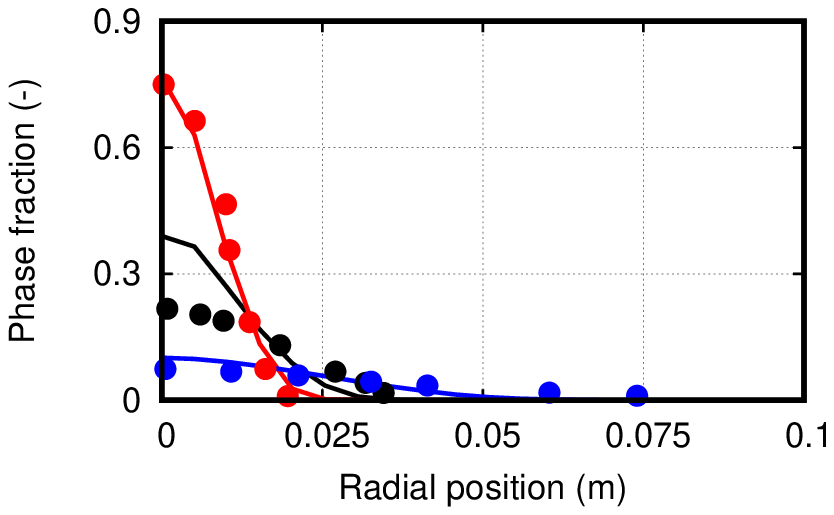}
\includegraphics
[width=0.48\textwidth]{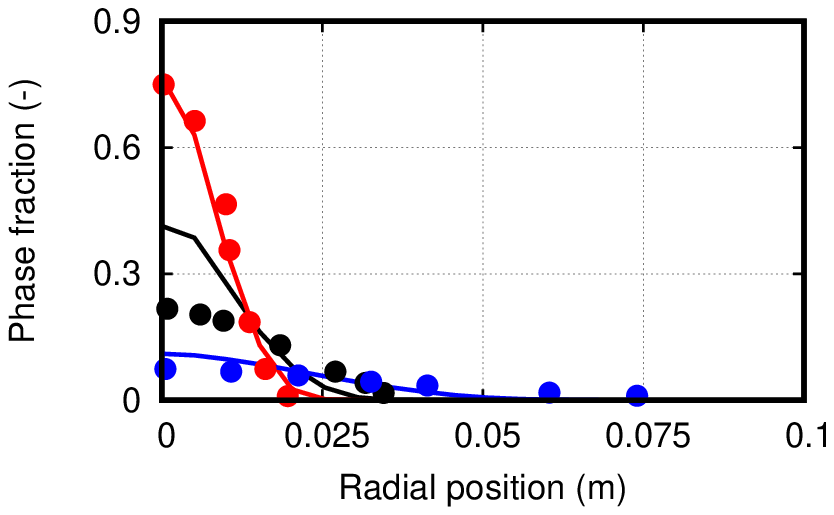}
\end{center}
\caption{Comparison of the phase fraction predicted by drag force and turbulence dispersion force (line) with experimental data (points) at different cross-sections of air-water
plumes. Red: $z=20$ mm. Black:  $z=100$ mm. Blue: $z=350$ mm. } 
\label{caseC2}
\end{figure}

Fig. \ref{caseC2sheng} shows the predicted axial liquid velocities, phase fraction and the turbulent kinetic energy. It can be seen that the predicted axial liquid velocity agree well with experiments if the drag force and turbulence dispersion force were included. Predictions became worse if the turbulence dispersion force was neglected. Including the lift and wall forces cannot improve the results, which implies their effects can be omitted. On the contrary with the test cases investigated previously, when only the drag force was included, the predicted turbulence kinetic energy was better than that predicted by all forces combination. However, the predicted phase fraction was seriously over-estimated, which implies the bubbles were not diffused and accumulated at the center line of the reactor. Fig. \ref{caseC2} shows the predicted phase fraction compared with experiments investigated by Castillejos et al. usting a similar equipment \cite{castillejos1987measurement}. The bubble flow rate is 257 N cm$^3$/s. It can be seen that the predicted results agree well with the experiments if the drag force and turbulence dispersion force were included.

%
%
%
%

\section{Conclusions} 

In this work, we employed different force combinations to simulate industrial bubbly flows. Instead of investigating the effects of momentum interfacial exchange force by a specific test case, nine different test cases were employed. Our aim is to seek a general numerical settings for the Eulerian-Eulerian model, which can provide competent results for industrial bubbly flow simulations. These test cases were selected from different industries including chemical, nuclear, bio-processing and metallurgical engineering.  Simulations were launched by the OpenFOAM solver \verb+reactingTwoPhaseEulerFoam+. The drag force developed by Ishii and Zuber \cite{ishii1979drag}, turbulent dispersion force developed by Lopez De Bertodano \cite{lopez1993turbulent}, lift force developed by Tomiyama et al. \cite{tomiyama2002transverse} and wall force developed by Antal et al. \cite{antal1991analysis} were employed.  Predictions were compared against experimental data. It was found that the drag force and turbulent dispersion force play the most important role on the predictions and should be included for all simulations. Otherwise the bubbles tend to accumulate since the spreading effect of the lift force is weak. For the pipe flows with large aspect ratio, the lift force and wall lubrication force should be included to address the phase fraction accumulation in the vicinity of the wall. In other cases these lateral forces can be safely neglected. 

At last, the test case library is open-sourced and are available as supplementary data for anyone to download as baseline test cases for further investigations.

\section*{Contributions} 
Yulong Li: wrote the paper; post-processed   the data; performed the analysis. Dongyue Li:  conceived the analysis; corrected the paper; prepared the OpenFOAM test cases.

\section*{Acknowledgement} 
One of the author (Dongyue Li) wants to acknowledge the CFD-China community for the fruitful discussions of gas-liquid flows. 

\section*{Appendix}

The finite volume discretization for the E-E method was implemented  in OpenFOAM-6 by the OpenFOAM foundation with novel  techniques to ensure stabilities. To the authors' knowledge, these techniques were not published. Therefore, we summarize the procedures in this section. After omitting the buoyancy and pressure terms, Eq.~(\ref{contAlpha}) and (\ref{contAlpha2}) can be written as follows: 
\begin{equation}\label{momentum3}
\frac{{\p  {{\alpha_\ra }{\rho_\ra }{\bfU_\ra }} }}{{\p t}} + \nabla \cdot \left( {{\alpha_\ra}{\rho_\ra } {{\bfU_\ra}  {\bfU_\ra}} } \right) - \\ \nabla  \cdot \left( {{\alpha_\ra}\rho_\ra{\bftau_\ra}} \right) 
= 
- {\bfK \bfd \bfU_\ra} - {\bfM_{\rm lift}} - {\bfM_{\rm turb}} - {\bfM_{\rm wall}},
\end{equation}
\begin{equation}\label{momentum4}
\frac{{\p  {{\alpha_\rb}{\rho_\rb}{\bfU_\rb}} }}{{\p t}} + \nabla \cdot \left( {{\alpha_\rb}{\rho_\rb} {{\bfU_\rb}  {\bfU_\rb}} } \right) - \\ \nabla  \cdot \left( {{\alpha_\rb}\rho_\rb{\bftau_\rb}} \right)  
= 
- {\bfK \bfd \bfU_\rb} + {\bfM_{\rm lift}} + {\bfM_{\rm turb}} + {\bfM_{\rm wall}},
\end{equation}
where
\begin{equation}\label{kd}
{\bfK \bfd} = \frac{3}{4} \alpha_\rb \rho_\rb {\rm C_{\rm D}}  \frac{1}{d_{\rm b}} |\bfU_{\rm a} - \bfU_{\rm b}|.
\end{equation}
The discretized form of Eq.~(\ref{momentum3}) and (\ref{momentum4}) can be written as follows:
\begin{equation}\label{FVM1}
A_{\rm a,P}{\bfU_{\rm a,P}} + \sum A_{\rm a,N}{\bfU_{\rm a,N}} = S_{\rm a,P},
\end{equation}
\begin{equation}\label{FVM2}
A_{\rm b,P}{\bfU_{\rm b,P}} + \sum A_{\rm b,N}{\bfU_{\rm b,N}} = S_{\rm b,P},
\end{equation}
where the subscript $_\rP$ denotes the owner cell, the subscript $_\rN$ denotes the neighbour cells, $A_{\rm P}$ and $A_{\rm N}$ are  the matrix diagonal coefficients contributed by the owner cells and neighbour cells, $S$ are source terms. The predicted velocities can be obtained by solving the discretized form of Eq.~(\ref{FVM1}) and (\ref{FVM2}):
\begin{equation}\label{predvelo1}
{\HbyA_{\ra,\rP}} = \frac{1}{A_{\rm a,P}}(-\sum A_{\rm a,N}{\bfU_{\rm a,N}} + S_{\rm a,P}),
\end{equation}
\begin{equation}\label{predvelo2}
{\HbyA_{\rb,\rP}} = \frac{1}{A_{\rm b,P}}(-\sum A_{\rm b,N}{\bfU_{\rm b,N}} + S_{\rm b,P}),
\end{equation}
where ${\HbyA_{\ra,\rP}}$ and ${\HbyA_{\rb,\rP}}$ are the predicted velocities for phase a and phase b, respectively. At this stage, the predicted velocity fields are not divergence free and a pressure equation needs to be constructed. 
In order to simplify the designation of the pressure boundary conditions and to reduce the spurious currents  caused by hydrostatic pressure in the non-orthogonal grids, the pressure term and buoyancy term are usually combined together, and the pressure without the hydrostatic part, $p_{\rm rgh} = p - \rm \rho \bfg \cdot \bfh$, is used. The gradient of $p_{\rm rgh}$ can be calculated by
\begin{equation}\label{prgh2}
\nabla p_{\rm rgh} = \nabla p - \rho \bfg - \bfg \cdot \bfh \nabla \rm \rho,
\end{equation}
Substituting the definaiton of $p_{\rm rgh} $ and \eq{prgh2} into the buoyance and pressure terms leads to:
\begin{multline}\label{fvm3}
-\alpha_{\rm a} \nabla p + \alpha_{\rm a} \rho_{\rm a} \bfg = -\alpha_{\rm a} \nabla p_{\rm rgh} - \alpha_{\rm a} \rho \bfg - \alpha_{\rm a} \bfg \cdot \bfh \nabla \rho + \alpha_{\rm a} \rho_{\rm a} \bfg 
\\= \alpha_{\rm a} \nabla p_{\rm rgh} - \alpha_{\rm b} \alpha_{\rm a} (\rho_{\rm b} - \rho_{\rm a}) \bfg - \alpha_{\rm a} \bfg \cdot \bfh \nabla \rm \rho,
\end{multline}
\begin{multline}\label{fvm4}
-\alpha_{\rm b} \nabla p + \alpha_{\rm b} \rho_{\rm b} \bfg = -\alpha_{\rm b} \nabla p_{\rm rgh} - \alpha_{\rm b} \rho \bfg - \alpha_{\rm b} \bfg \cdot \bfh \nabla \rho + \alpha_{\rm b} \rho_{\rm b} \bfg 
\\ = \alpha_\rb \nabla p_\rgh - \alpha_{\rm a} \alpha_{\rm b} (\rho_{\rm a} - \rho_{\rm b}) \bfg - \alpha_{\rm b} \bfg \cdot \bfh \nabla \rm \rho,
\end{multline}
In this manner, a $p_{\rm rgh}$ equation can be constructed.

For incompressible flows, summarizing \eq{contAlpha} and \eq{contAlpha2} produces the divergence constraint equation:
\begin{equation}\label{sumdensity}
\nabla \cdot (\alpha_{\rm b} \bfU_{\rm b} + \alpha_{\rm a} \bfU_{\rm a})= 0.
\end{equation}
The discretized form of \eq{sumdensity} can be written as follows: 
\begin{equation}\label{restrcondi}
\sum(\alpha_{\rm a,f} \bfU_{\rm a,f} + \alpha_{\rm b,f} \bfU_{\rm b,f}) \cdot \bfS_{\rm f} = 0.
\end{equation}
where the subscript $_\mathrm{f}$ implies variables defined at the cell faces.  Eq.~(\ref{restrcondi}) is usually employed as a restrictive condition to construct pressure Poisson equation. Since the pressure gradient and buoyancy term are not considered in Eq.~(\ref{predvelo1}) and (\ref{predvelo2}), adding these terms produces the predicted velocities:
\begin{equation}\label{Uap}
\bfU_{\rm a,P} = \bfH \bfb \bfy\bfA_{\rm a,P} + \frac{\alpha_{\rm a,P}}{\bfA_{\rm a,P}}(\nabla p_{\rm rgh,P} - \alpha_{\rm b,P}(\rho_{\rm b} - \rho_{\rm a})\bfg - \bfg \cdot \bfh_{\rm P} \nabla \rho_\rP) + \frac{\rm Kd_P}{\bfA_{\rm a,P}}\bfU_{\rm b,P},
\end{equation}
\begin{equation}\label{Ubp}
\bfU_{\rm b,P} = \bfH \bfb \bfy\bfA_{\rm b,P} + \frac{\alpha_{\rm b,P}}{\bfA_{\rm b,P}}(\nabla p_{\rm rgh,P} - \alpha_{\rm a,P}(\rho_{\rm a} - \rho_{\rm b})\bfg - \bfg \cdot \bfh_{\rm P} \nabla \rho_\rP) + \frac{\rm Kd_P}{\bfA_{\rm b,P}}\bfU_{\rm a,P}.
\end{equation}
Substituting the interpolated cell face velocities into Eq.~(\ref{restrcondi}), the phase fluxes can be written as follows: 
\begin{equation}\label{FVM5}
\alpha_{\rm b,f}\phi_{\rm b,f} + \alpha_{\rm a,f}\phi_{\rm a,f} = \nabla \cdot \Bigg( \Big(\alpha_{\rm b,P}\frac{\alpha_{\rm b,P}}{\bfA_{\rm b,P}} + \alpha_{\rm a,P}\frac{\alpha_{\rm a,P}}{\bfA_{\rm a,P}} \Big) \nabla p_{\rm rgh,P} \Bigg),
\end{equation}
where
\begin{equation}\label{phia}
\phi_{\rm a,f} = (\bfH \bfb \bfy\bfA_{\rm a,f} + \frac{\alpha_{\rm a,f}}{\bfA_{\rm a,f}}(-\alpha_{\rm b,f}(\rho_{\rm b} - \rho_{\rm a})\bfg - \bfg \cdot \bfh_{\rm f} \nabla \rm \rho_{\rm f}) + \frac{\rm Kd_f}{\bfA_{\rm a,f}}\bfU_{\rm b,f}) \cdot \bfS_{\rm f},
\end{equation}
\begin{equation}\label{phib}
\phi_{\rm b,f} = (\bfH \bfb \bfy\bfA_{\rm b,f} + \frac{\alpha_{\rm b,f}}{\bfA_{\rm b,f}}(-\alpha_{\rm a,f}(\rho_{\rm a} - \rho_{\rm b})\bfg - \bfg \cdot \bfh_{\rm f} \nabla \rm \rho_{\rm f}) + \frac{\rm Kd_f}{\bfA_{\rm b,f}}\bfU_{\rm a,f}) \cdot \bfS_{\rm f}.
\end{equation}
Eq.~(\ref{FVM5}) can be used to calculate the $p_\mathrm{rgh}$ in the PISO loops \cite{issa1986computation}. After the pressure was updated, a divergence-free velocities can be updated. 

When the drag coefficient $\bfK\bfd$ is much larger than the diagonal coefficient $\bfA$ at certain cells, the classical semi-implicit algorithm discussed previously leads to very large relative velocity Courant number and the time step will be very small. In OpenFOAM-6, a robust approach was implemented by Weller \cite{weller2019} to handle the pressure-velocity coupling and it was not published.  In this method, the velocities without contributions of the drag forces are defined as follows:         
\begin{equation}\label{Uap2}
\bfU_{\rm a,P}^{\rm s} = \bfH \bfb \bfy\bfA_{\rm a,P} + \frac{\alpha_{\rm a,P}}{\bfA_{\rm a,P}}(\nabla p_{\rm rgh,P} - \alpha_{\rm b,P}(\rho_{\rm b} - \rho_{\rm a})\bfg - \bfg \cdot \bfh_{\rm P} \nabla \rho_{\rm P}),
\end{equation}
\begin{equation}\label{Ubp2}
\bfU_{\rm b,P}^{\rm s} = \bfH \bfb \bfy\bfA_{\rm b,P} + \frac{\alpha_{\rm b,P}}{\bfA_{\rm b,P}}(\nabla p_{\rm rgh,P} - \alpha_{\rm a,P}(\rho_{\rm a} - \rho_{\rm b})\bfg - \bfg \cdot \bfh_{\rm P} \nabla \rho_{\rm P}).
\end{equation}
The phase velocities can be written as
\begin{equation}\label{Uap3}
\bfU_{\rm a,P} = \bfU_{\rm a,P}^{\rm s} + \rm D_{\rm a,P}\bfU_{\rm b,P},
\end{equation}
\begin{equation}\label{Ubp3}
\bfU_{\rm b,P} = \bfU_{\rm b,P}^{\rm s} + \rm D_{\rm b,P}\bfU_{\rm a,P},
\end{equation}
\begin{equation}\label{Up}
\bfU_{\rm P} = \alpha_{\rm b,P}( \bfU_{\rm b,P}^s + \rm D_{\rm b,P}\bfU_{\rm b,P} ) + \alpha_{\rm a,P}( \bfU_{\rm a,P}^s + \rm D_{\rm a,P}\bfU_{\rm b,P} ),
\end{equation}
where $\rm D_{\rm b,P} = \frac{\rm Kd_p}{\bfA_{\rm b,p}}$, $\rm D_{\rm a,P} = \frac{\rm Kd_p}{\bfA_{\rm a,p}}$. The relative velocity is defined by
\begin{equation}\label{Upr}
\bfU_{\rm P}^{\rm r} = \bfU_{\rm b,P} - \bfU_{\rm a,P} =  \bfU_{\rm b,P}^s + \rm D_{\rm b,P}\bfU_{\rm a,P} - \bfU_{\rm a,P}^s - \rm D_{\rm a,P}\bfU_{\rm b,P}.
\end{equation}
Subtracting $\rm D_{\rm b,P}\rm D_{\rm a,P}(\bfU_{\rm b,P} - \bfU_{\rm a,P})$ from the L.H.S. and the R.H.S. of Eq.~(\ref{Upr}) leads to: 
\begin{multline}\label{Upr1}
\bfU_{\rm P}^{\rm r}- \rm D_{\rm b,P}\rm D_{\rm a,P}(\bfU_{\rm b,P} - \bfU_{\rm a,P}) =\\ \bfU_{\rm b,P}^s + \rm D_{\rm b,P}\bfU_{\rm a,P} - \bfU_{\rm a,P}^s - \rm D_{\rm a,P}\bfU_{\rm b,P} - \rm D_{\rm b,P}\rm D_{\rm a,P}(\bfU_{\rm b,P} - \bfU_{\rm a,P}).
\end{multline}
After arrangement, \eq{Upr1} can be written as follows:
\begin{equation}\label{Upr2}
\bfU_{\rm P}^{\rm r} = \frac{(1 - \rm D_{\rm a,P})\bfU_{\rm b,P}^s - (1 - \rm D_{\rm b,P})\bfU_{\rm a,P}^s}{1 - \rm D_{\rm a,P}\rm D_{\rm b,P} }.
\end{equation}
Once the relative velocity $\bfU_{\rm P}^{\rm r}$ is updated, the velocity of discrete phase and continuous phase can be calculated by the following equations, respectively: 
\begin{equation}\label{}
\bfU_{\rm a,P} = \bfU_{\rm P} - \alpha_{\rm b,P}\bfU_{\rm P}^{\rm r},
\end{equation}
\begin{equation}\label{}
\bfU_{\rm b,P} = \bfU_{\rm P} + \alpha_{\rm a,P}\bfU_{\rm P}^{\rm r}.
\end{equation}

For the flows with strong swirl or strong body forces, the algorithm discussed above tends to  produce phase fraction oscillations \cite{zhang2014generalized}. 
This issue can be mitigated by including these body forces on the cell faces to construct the face-based pressure equation. In this manner,  \eq{momentum3} and \eq{momentum4} can be re-written as follows:
\begin{equation}\label{momentum5}
\frac{{\p \left( {{\alpha_\ra }{\rho_\ra }{\bfU_\ra }} \right)}}{{\p t}} + \nabla \cdot \left( {{\alpha_\ra}{\rho_\ra } {{\bfU_\ra}  {\bfU_\ra}}} \right) - \\ \nabla  \cdot \left( {{\alpha_\ra}\rho_\ra{\bftau_\ra}} \right) 
= 
- {\bfK \bfd \bfU_\ra} ,
\end{equation}
\begin{equation}\label{momentum6}
\frac{{\p \left( {{\alpha_\rb}{\rho_\rb}{\bfU_\rb}} \right)}}{{\p t}} + \nabla \cdot \left( {{\alpha_\rb}{\rho_\rb} {{\bfU_\rb}  {\bfU_\rb}} } \right) - \\ \nabla  \cdot \left( {{\alpha_\rb}\rho_\rb{\bftau_\rb}} \right)  
= 
- {\bfK \bfd \bfU_\rb}.
\end{equation}
It can be seen that in \eq{momentum5} and \eq{momentum6} the contribution of the non-drag forces is neglected. The contribution needs to be considered to update the predicted velocities. Therefore, \eq{Uap} and \eq{Ubp} can be re-written as follows:
\begin{multline}\label{Uap2}
\bfU_{\rm a,p} = \bfH \bfb \bfy\bfA_{\rm a,P} + \frac{\alpha_{\rm a,P}}{\bfA_{\rm a,P}}(\nabla p_{\rm rgh,P} - \alpha_{\rm b,P}(\rho_{\rm b} - \rho_{\rm a})\bfg - \bfg \cdot \bfh_{\rm P} \nabla \rho_\rP)\\ +\frac{\rm Kd_P}{\bfA_{\rm a,P}}\bfU_{\rm b,P}- {\bfM_{\rm lift,P}} - {\bfM_{\rm turb,P}} - {\bfM_{\rm wall,P}},
\end{multline}
\begin{multline}\label{Ubp3}
\bfU_{\rm b,p} = \bfH \bfb \bfy\bfA_{\rm b,P} + \frac{\alpha_{\rm b,P}}{\bfA_{\rm b,P}}(\nabla p_{\rm rgh,P} - \alpha_{\rm a,P}(\rho_{\rm a} - \rho_{\rm b})\bfg - \bfg \cdot \bfh_{\rm P} \nabla \rho_\rP)\\ + \frac{\rm Kd_P}{\bfA_{\rm b,P}}\bfU_{\rm a,P}+{\bfM_{\rm lift,P}}+ {\bfM_{\rm turb,P}} + {\bfM_{\rm wall,P}}.
\end{multline}
The corresponding fluxes can be written as follows:
\begin{multline}\label{phia2}
\phi_{\rm a,f} = \left(\bfH \bfb \bfy\bfA_{\rm a,f} + \frac{\alpha_{\rm a,f}}{\bfA_{\rm a,f}}(-\alpha_{\rm b,f}(\rho_{\rm b} - \rho_{\rm a})\bfg - \bfg \cdot \bfh_{\rm f} \nabla \rm \rho_{\rm f}) + \frac{\rm Kd_f}{\bfA_{\rm a,f}}\bfU_{\rm b,f}\right) \cdot \bfS_{\rm f}
\\
-\left({\bfM_{\rm lift,P}}+ {\bfM_{\rm turb,P}} + {\bfM_{\rm wall,P}}\right)\cdot \bfS_{\rm f},
\end{multline}
\begin{multline}\label{phib2}
\phi_{\rm b,f} = \left(\bfH \bfb \bfy\bfA_{\rm b,f} + \frac{\alpha_{\rm b,f}}{\bfA_{\rm b,f}}(-\alpha_{\rm a,f}(\rho_{\rm a} - \rho_{\rm b})\bfg - \bfg \cdot \bfh_{\rm f} \nabla \rm \rho_{\rm f}) + \frac{\rm Kd_f}{\bfA_{\rm b,f}}\bfU_{\rm a,f}\right) \cdot \bfS_{\rm f}
\\
+\left({\bfM_{\rm lift,P}}+ {\bfM_{\rm turb,P}} + {\bfM_{\rm wall,P}}\right)\cdot \bfS_{\rm f}.
\end{multline}
Substituting $\phi_{\rm a,f}$ and $\phi_{\rm b,f}$ into \eq{FVM5} leads to the pressure Poisson equation which can avoid the oscillating results due to strong body forces. 

\section*{References}
\bibliographystyle{unsrtnat}
\bibliography{manuscript}

\end{document}